\documentclass[a4paper,fleqn,usenatbib,useAMS]{mnras}
\usepackage{upgreek}
\usepackage{graphicx}
\usepackage{amsmath}
\usepackage{amssymb}
\usepackage{bm}
\usepackage{pdflscape}
\usepackage{float}
\usepackage[T1]{fontenc}
\usepackage{ae,aecompl}
\usepackage{color}
\usepackage{caption}
\usepackage[shortlabels]{enumitem}

\pdfoutput=1

\title[Tidal Dissipation in Massive Binary Star Systems]{Tidal Dissipation Impact on the Eccentric Onset of Common Envelope Phases in Massive Binary Star Systems}

\author[M. Vick et al.]{Michelle Vick$^{1}$, Morgan MacLeod$^{2}$, Dong Lai$^{3}$, and Abraham Loeb$^{2}$\\$^{1}$Center for Interdisciplinary Exploration \& Research in Astrophysics (CIERA), Northwestern
	University, Evanston, IL 60208, USA
\\$^{2}$Center for Astrophysics $\vert$ Harvard \& Smithsonian, 60 Garden Street, Cambridge, MA, 02138, USA
\\$^{3}$Cornell Center for Astrophysics and Planetary Science, Department of Astronomy, Cornell University, Ithaca, NY 14853, USA}

\pubyear{2020}

\hypersetup{draft}
\begin{document}
	
% Make the title.
	
\label{firstpage}
\pagerange{\pageref{firstpage}--\pageref{lastpage}}
\maketitle
	
\begin{abstract}
Tidal dissipation due to turbulent viscosity in the convective regions of giant stars plays an important role in shaping the orbits of pre-common envelope systems. Such systems are possible sources of transients and close compact binary systems that will eventually merge and produce detectable gravitational wave signals. Most previous studies of the onset of common envelope episodes have focused on circular orbits and synchronously rotating donor stars under the assumption that tidal dissipation can quickly spin up the primary and circularize the orbit before the binary reaches Roche-lobe overflow (RLO). We test this assumption by coupling numerical models of the post main sequence stellar evolution of massive-stars with the model for tidal dissipation in convective envelopes developed in \citet{Vick20} --- a tidal model that is accurate even for highly eccentric orbits with small pericentre distances. We find that, in many cases, tidal dissipation does not circularize the orbit before RLO. For a $10~M_\odot$ ($15~M_\odot$) primary star interacting with a $1.4~M_\odot$ companion, systems with pericentre distances within 3~AU (6~AU) when the primary leaves the main sequence will retain the initial orbital eccentricity when the primary grows to the Roche radius. Even in systems that tidally circularize before RLO, the donor star may be rotating subsynchronously at the onset of mass transfer. Our results demonstrate that some possible precursors to double neutron star systems are likely eccentric at the Roche radius. The effects of pre-common envelope eccentricity on the resulting compact binary merit further study.
\end{abstract}
	

\section{Introduction}\label{sec:Intro}	

Common envelope phases occur in binary systems when one star evolves and grows in radius such that it impinges on the orbit of its companion. As the two stellar cores become subsumed within a shared envelope, they spiral closer under the influence of drag forces \citep{1976IAUS...73...75P}. Thus, common envelope phases represent a brief but transformative episode in the evolution of many binary or multiple stellar systems \citep[for reviews, see][]{1993PASP..105.1373I,2000ARA&A..38..113T,2013A&ARv..21...59I,2017PASA...34....1D}. In transforming wide binaries into much more compact ones, common envelope phases are thought to be a crucial element in the assembly of compact binaries that merge and produce gravitational wave sources \citep[e.g.][]{2000ARA&A..38..113T,2002ApJ...572..407B,2007PhR...442...75K,2008ApJS..174..223B,2012ApJ...759...52D,2013A&ARv..21...59I}. 

As recent work has focused on the common envelope phases that may lead to the formation of merging compact object binaries, it has become clear that some of the details of massive-star common envelope phases may be different from those in the previously-emphasized lower-mass systems \citep[as described in the reviews of][]{2000ARA&A..38..113T,2013A&ARv..21...59I}. In particular \citet{2016A&A...596A..58K} and \citet{2020arXiv200400628K} have studied pre-common envelope tar structures, with a focus on determining the binding energy of their envelopes, and \citet{2020arXiv200109829V} have performed a binary population synthesis study of all the common envelope phases that lead to merging double neutron star systems within the COMPAS suite \citep{2017NatCo...814906S,2018MNRAS.481.4009V}. 

This focus on massive-star evolution in close binaries has revealed the significance of tidal interactions in shaping these objects \citep[e.g.][]{2016MNRAS.462..844K}. In particular, in pre-common envelope systems, as the giant star evolves and grows in radius, there will be a period of time during which strong tides are active followed by eventual Roche-lobe overflow (RLO). In the massive-star progenitors of merging compact objects, it is not clear that tides will have sufficient time to act to always synchronize and circularize a system to the classic Roche-lobe geometry prior to RLO \citep{2020arXiv200109829V}. This paper applies new, more sophisticated models of tidal dissipation in giant star convective envelopes to study the impact of pre-common envelope tidal evolution. Our results, therefore, are of critical importance, and set the initial conditions for the subsequent common envelope or mass-transferring interaction.

There are two tidal dissipation mechanisms that can be important in a massive giant star with a radiative core and convective envelope. The first is turbulent viscosity, where the shearing of tidally driven fluid oscillations (fundamental and inertial modes) in the envelope is dissipated as heat \citep[e.g.][]{Zahn77, Goodman97,Ogilvie07}. The other is radiative dissipation in the stellar interior \citep[e.g.][]{Goodman98,Savonije02}. This second mechanism is important when gravity waves are excited at the radiative-convective boundary and grow in amplitude as they propagate inward. Eventually non-linear wave-braking dissipates all of the energy and angular momentum carried by the in-going gravity wave \citep{Goodman98,Barker10,Barker11,Chernov13,Ivanov13,Mathis16,Weinberg17,Sun18}. Even when gravity waves do not become non-linear, radiative damping can cause significant dissipation if the spectrum of g-modes in the core is very dense, and waves that are excited at the radiative-convective boundary damp before reaching the centre \citep{Ivanov13,Chernov17}. In the absence of these effects, dissipation in the convective envelope drives the orbital and spin evolution.

A large body of work has focused on orbital circularization due to tidal dissipation in the convective envelope of a star of a close binary. An analytical treatment for tidal dissipation via turbulent viscosity was first developed by \citet{Zahn77}. \citet{Verbunt95} presented a similar expression for the tidal circularization timescale for nearly circular binaries. The weak friction approximation \citep{Hut81} can be used more broadly to study the tidal evolution of binaries with any eccentricity \citep[as in][]{Hurley02}. This model assumes that the star is tidally deformed into a static shape that lags behind the equipotential surface --- an approximation that holds when the tidal forcing frequency, $\omega$, is much lower than the dynamical frequency of the star $\sim ({GM_1/R_1^3})^{-1/2}$ (for a star with mass $M_1$ and radius $R_1$). 

The weak friction model also assumes that turbulent viscosity is equally effective at dissipating energy for all tidal forcing frequencies. In reality, the turbulent viscosity is reduced when the timescale for tidal forcing $\sim \omega^{-1}$ is shorter than the turnover time for the largest convective eddies in the star $\tau_{\rm eddy}$. Two conflicting scaling laws have been suggested for the form of the viscosity reduction. \citet{Zahn89} proposed the linear reduction $1/(1+\omega \tau_{\rm eddy})$, while \citet{Goldreich77} favored $1/[1+(\omega \tau_{\rm eddy})^{2}]$. The latter is the standard result for a damped harmonic oscillator with frequency $\omega$ and damping time $\tau_{\rm eddy}.$ \cite{IP04b} used a normal mode decomposition to explore the effects of viscosity reduction with the general form $1/[1+(\omega \tau_{\rm eddy})^p],$ and found that when $p >1$, the binary could evolve through multiple resonances between the stellar rotation period and the orbital period. Recently, numerical and analytical studies have generally supported a quadratic reduction factor \citep{Penev11a,Penev11b,Ogilvie12,Duguid19}, although numerical simulations from \citet{Vidal20} and \citet{Duguid20} suggest that both scaling relations may be correct in different regimes of the tidal forcing frequency.

\citet{Vick20}, hereafter VL20, presented a general formalism for tidal evolution in an eccentric binary. The theory can be applied to various types of binaries where viscous dissipation in the convection zone dominates the tidal evolution. The formalism uses a normal mode decomposition, including the effect of stellar rotation \citep[see also][who adopted an approximate treatment of rotation in mode decomposition]{IP04b}. It allows for general tidal frequencies (compared to the dynamical frequency of the star) and accounts for frequency dependent damping of tidally-driven oscillations.
Although, for a giant star in a close binary, viscosity reduction typically is not important. VL20 shows that the rate of tidal circularization can be orders of magnitude faster than predicted by the weak friction approximation for highly eccentric binaries where the ratio of the pericentre distance $r_{\rm p}$ to stellar radius is of order a few.

In this paper, we study the coupled roles of stellar evolution and tidal dissipation in shaping pre-common envelope systems by applying the tidal theory of VL20. In Section \ref{sec:BigTheory}, we introduce the tools that we use to tackle this problem --- MESA generated stellar evolution models \citep{Paxton11} and the formalism from VL20 --- and discuss how we couple the two. In Section \ref{sec:Results}, we present and categorize the outcomes of concurrent stellar evolution and tidal orbital circularization. We focus on the orbital eccentricity and the primary star rotation rate at the onset of RLO. We also explore how tidal dissipation affects the distribution of those properties at the onset of binary mass transfer given an initial distribution of orbital parameters when the primary star leaves the main sequence. We discuss the limitations and significance of our results in Section~\ref{sec:Discussion}, and we summarize and conclude in Section~\ref{sec:Summary}.

\section{Theory of Tidal Dissipation in Eccentric Binaries} \label{sec:BigTheory}

In order to understand the coupled roles of stellar evolution and tidal dissipation we need: (i) a time-dependent model of the structure of the massive-star after it has left the main sequence; (ii) an understanding of how the timescale for tidal dissipation changes with the stellar structure; and (iii) a framework for calculating the tidal energy and angular momentum transfer rates given the current orbit and tidal dissipation rate. 

The weak friction model \citep{Alexander73,Hut81} is commonly used to describe equilibrium tidal interactions. In this model, the gravitational potential of a companion raises a tidal bulge on the primary star. When the binary orbit is eccentric or the primary is not synchronously rotating, the bulge lags behind the axis connecting the two bodies. The magnitude of the lag angle depends on the rate at which the star can dissipate the energy in tidally excited fluid motion. The weak friction theory assumes that the lag angle is the tidal frequency multiplied by a constant lag time. This constant-lag-time assumption is usually not correct.

When the primary star evolves off of the main sequence, the binary separation is large compared to the stellar radius, and tidal interactions are weak. However, as the stellar radius expands, the primary experiences a stronger tidal potential from its companion. Additionally, the timescale for strong tidal interactions in the binary (roughly the duration of a pericentre passage) begins to approach the dynamical time $(R_1^3/GM_1)^{1/2}$ of the giant star. Under these conditions, the standard weak friction model of tides breaks down, and can severely underestimate the strength of tidal interactions. We use the theory of dissipation in a giant star in an eccentric orbit developed in VL20 to study how stellar evolution and tidal dissipation jointly shape the orbital evolution of the binary.

\subsection{Stellar Models} \label{sec:StellarModels}

We have used version 11701 of the MESA stellar evolution code to calculate the structural evolution of $10 M_\odot$ and $15 M_\odot$ stars from the end of the main sequence to carbon depletion \citep{Paxton11}. We assumed an initial metallicity of $Z=Z_\odot = 0.0142$ \citep{Asplund09}, and used the ``Dutch" wind scheme. The inlist to reproduce our calculations will be made available at the MESA Marketplace (http://cococubed.asu.edu/mesa\_market/inlists.html). 
The result is a suite of stellar profiles at different time stamps in the star's evolution. The time interval between profiles ranges from 10 years (during periods of rapid radius expansion) to $5\times 10^5$ years when the stellar structure is relatively static.

At each timestep, we used the code GYRE to calculate the eigenfrequency and mode profile of the stellar $l=2$ f-mode for a given MESA model (assuming no rotation) \citep{Townsend13}. We implemented a vacuum outer boundary and a zero radial displacement inner boundary. For models with a convective envelope, the transition between the core and envelope was used as the location of the inner boundary. We used the condition on the convective velocity $v_{\rm c} (r) > 10^3$~cm/s to identify $r_{\rm c}$, the start of the envelope. Note that, after a deep convective envelope has developed, $r_{\rm c}/R_1 \ll 1$. For models from earlier in the star's evolution, before the development of deep convection, we chose an inner boundary just outside of the composition transition from predominantly hydrogen to predominantly helium. 

\subsection{Calculation of the Tidal Dissipation Rate in a Giant Star}\label{sec:Models}

As the primary $M_1$ transitions to core helium burning, the star develops a deep convective envelope. Within this outer region, turbulent viscosity can dissipate tidally excited fluid motion, sapping energy and angular momentum from the orbit. 

\subsubsection{Order of Magnitude Calculation}\label{sec:OrderofMag}
A simple estimate of the tidal circularization time for a nearly-circular binary is provided in, e.g. \cite{Zahn77,Phinney92}; and \cite{Verbunt95}. Adopting a typical value $\nu_0$ for the viscosity in the convective envelope, the damping rate of a tidally forced oscillation is,
\begin{equation}
\gamma_{\rm est} \sim \frac{M_{\rm env}}{M_1}\left(\frac{\nu_0}{H^2}\right) \sim \frac{M_{\rm env}}{M_1}\left(\frac{L}{M_{\rm env}R_1^2}\right)^{1/3}, 
\label{eq:gamma_est}
\end{equation}
where $H$ is the pressure scale-height (and the length-scale of the largest convective eddies), $M_{\rm env}$ is the mass of the envelope, and $L$ is the convective luminosity. We have used $\nu_0\sim H(L/4\uppi\rho R_1^2)^{1/3}$ and $M_{\rm env} \sim 4 \uppi \rho H^3$, where $\rho$ is the average density in the convective envelope. For a nearly circular orbit, this damping rate is related to the binary circularization time $\tau_{\rm circ}$ via,
\begin{equation}
t_{\rm circ} \equiv \left|\frac{e}{\dot{e}}\right| \sim \frac{1}{\gamma_{\rm est}}\left(\frac{M_1}{M_2}\right)\left(\frac{M_1}{M_t}\right)\left(\frac{a}{R_1}\right)^8,\label{eq:tauCircSimple}
\end{equation}
with $M_2$ the mass of the companion, $M_t = M_1+M_2$ and semi-major axis $a$ \citep[see][]{Phinney92}.

With a given stellar profile, we can calculate the damping rate $\gamma_{\rm f}$ for a forced f-mode oscillation more precisely. The response of $M_1$ to the tidal potential of $M_2$ is dominated by the quadrupolar $l=2$ terms (if the binary is sufficiently separated). In general, the damping rate $\gamma_{\rm f}$ depends on the tidal forcing frequency $\omega$ [see equation (19) of VL20]. When the turnover time for the largest convective eddies is shorter than the timescale for tidal forcing, the viscosity in the envelope is not reduced. In a red giant star, the eddy turnover time in the convective envelope is typically short relative to the tidal forcing period, and this condition is satisfied while $\omega \ll (GM_1R_1^{-3})^{1/2}$ (see the top panel of Fig. 2 in VL20).

\subsubsection{Weak Friction Approximation and Nearly Circular Orbits}\label{sec:WeakFriction}

When the viscous damping rate in the envelope is independent of the tidal forcing frequency, the tidal evolution equations can be framed in terms of the stellar tidal Love number and lag time. If the tidal forcing period is much longer than the dynamical time of the star, this treatment is equivalent to the weak friction approximation. For an $l=2$ f-mode oscillation in a slowly rotating
body, the real part of the tidal Love number is, 
\begin{equation}
k_2 \simeq \frac{4\uppi}{5}\left(\frac{Q_{\rm f}}{\bar{\omega}_{\rm f}}\right)^2, \label{eq:defk2}
\end{equation}
where $Q_{\rm f}$ is an overlap integral defined in equation (12) of VL20, normalized such that $G=M_1=R_1=1$, and $\omega_{\rm f} \equiv \bar{\omega}_{\rm f} (GM_1/R_1^3)^{1/2}$ is the f-mode frequency. The tidal lag time is given by, 
\begin{equation}
\tau \equiv \frac{\gamma_{\rm f}}{\omega_{\rm f}^2}, \label{eq:deftau}
\end{equation} 
with $\gamma_{\rm f}$ the damping rate of the $l=2$ f-mode due to turbulent viscosity [equation (19) of VL20].

Under the condition that the tidal forcing frequency $\omega \ll \omega_{\rm f}$, we can express the circularization rate for a synchronously rotating star in a nearly circular binary as, \citep{Darwin1880,Alexander73,Hut81}
\begin{equation}
\frac{\dot{e}}{e} = -\frac{21}{2}k_2 \tau \Omega^2 \frac{M_2}{M_1}\left(\frac{R_1}{a}\right)^5, \label{eq:tcircSimpleWF}
\end{equation}
where $\Omega = (GM_t/a^3)^{1/2}$ is the orbital frequency. By comparing equation~(\ref{eq:tcircSimpleWF}) to equation~(\ref{eq:tauCircSimple}),
we find that the effective damping rate from tidal dissipation is,
\begin{equation}
\gamma_{\rm eff} \equiv \frac{21}{2}k_2\tau\left(\frac{GM_1}{R_1^3}\right)\label{eq:gammaEff}.
\end{equation}

\subsubsection{Theory of Tides and Dissipation in an eccentric binary}\label{sec:Theory}
%A neutron star-giant star binary may be left with significant eccentricity form the supernova that formed the compact companion. 
VL20 developed a general formalism for the treatment of tidal dissipation in the convective envelope of a star in an eccentric binary. For an eccentric orbit, the quadrupolar tidal potential experienced by the primary star $M_1$ from the companion $M_2$ can be decomposed into a sum over many forcing frequencies.
% \begin{equation}
% \omega_{Nm} \equiv N\Omega - m \Omega_{\rm s}, \label{eq:UNm}
% \end{equation}
% where $N$ is an integer. In the rotating frame of the primary, the potential is given by
% \begin{equation}
% U(\boldsymbol{r}, t) = -  \sum_{m}\sum_{N=-\infty}^{\infty} U_{Nm} r^2 Y_{2m}(\theta,\phi)\text{e}^{-\text{i}\omega_{Nm}t},\label{eq:potentialSum}
% \end{equation} 
% where $\boldsymbol{r} = (r,\theta,\phi)$ is the position vector in spherical coordinates relative to the center of mass of the primary star, and the angle $\phi$ is measured in the rotating frame of $M_1$. Throughout the paper, we assume that the spin-axis of $M_1$ is aligned with the orbital angular momentum axis. We define
% \begin{align}
% U_{Nm} &\equiv \frac{GM_2}{a^3} W_{2m} F_{Nm}
% \end{align}
% with $F_{Nm}$ given by
% \begin{equation}
% F_{Nm} = \frac{1}{\uppi}\int_0^\uppi d\Psi \; \frac{\cos[N(\Psi - e \sin \Psi)-m\Phi(t)]}{(1-e\cos \Psi)^2}.
% \end{equation}
% Only the $m =  0, \pm2$ terms are nonzero, with $W_{20} = \sqrt{\uppi/5}$ and $W_{2\pm2} = \sqrt{3 \uppi/10}$.
The tidal response of $M_1$ is a weighted sum of the response to each frequency term in the tidal potential. Using this formalism, VL20 derived expressions for the tidal torque and energy transfer
rate (see their equations 43, 44, 51 and 52). We use their results in the orbital evolution equations presented in the following subsection (Section~\ref{sec:Methods}).

VL20 allowed for viscosity reduction when
the tidal forcing time is shorter than the turnover time for convective eddies in the primary star. For a giant star, the tidal forcing time is often longer than the eddy turnover time, and viscosity reduction is negligible. Thus $\gamma_{\rm f}$ is a constant for a given stellar model. In this case, VL20 also demonstrated that their treatment yields equivalent orbital evolution equations to the weak friction approximation [equations (9)-(11) of \citep{Hut81}] in the limit that $\omega \ll \omega_{\rm f}$. However, for highly eccentric orbits with small pericentre distances $r_{\rm p}$, the tidal forcing frequency $\omega$ can be comparable to $\omega_{\rm f}$. Under these conditions, VL20 found that the torque and energy transfer rate
can be orders of magnitude larger than the weak friction calculation suggests (see Fig. 5 of VL20).

\subsection{Coupling the Stellar and Orbital Evolution}\label{sec:Methods}
	In order to couple the stellar evolution and orbital evolution, we use the stellar oscillation code GYRE to calculate the properties of the f-mode (e.g. $\gamma_{\rm f}$, $\bar{\omega}_{\rm f}$, and $Q_{\rm f}$). We then use spline interpolation to obtain the stellar mass, radius, and mode properties as a function of time.
	
	The time evolution of $a$, $e$, and $\Omega_{\rm s}$  is given by the following equations:
	\begin{align}
	\frac{\dot{a}}{a} &= \left.\frac{\dot{a}}{a}\right|_{\rm Tides} + \left.\frac{\dot{a}}{a}\right|_{\rm Wind} \\
	\frac{\dot{\Omega}_{\rm s}}{\Omega_{\rm s}} &= \left.\frac{\dot{\Omega}_{\rm s}}{\Omega_{\rm s}}\right|_{\rm Tides}  - \frac{\dot{I}}{I}, \\
	\frac{\dot{e}}{e} &= \left.\frac{\dot{e}}{e}\right|_{\rm Tides},
	\end{align}
	where $I = \eta M_1 R_1^2$
	is the moment of inertia of the primary, and the time evolution of $\eta$, $M_1$, and $R_1$ is taken from the MESA stellar models.
	The tidal energy transfer and torque can be combined to determine the tidal contributions to the orbital evolution,
	\begin{align}
	\left.\frac{\dot{a}}{a}\right|_{\rm Tides} &= -\frac{6}{t_d(1-e^2)^{15/2}}F_E(e,\Omega_{\rm s}/\Omega,r_{\rm p}/R_1),\label{eq:Simple_adot}\\
	\left.\frac{\dot{\Omega}_s}{\Omega_{\rm s}}\right|_{\rm Tides} &= \frac{3}{t_d(1-e^2)^{6}}\left(\frac{\mu a^2 \Omega}{I \Omega_{\rm s}}\right)F_T(e,\Omega_{\rm s}/\Omega,r_{\rm p}/R_1), \label{eq:OmegaSdot}\\
	\left.\frac{\dot{e}}{e}\right|_{\rm Tides} &= -\frac{27}{t_d(1-e^2)^{13/2}} F_{\rm ecc}(e,\Omega_{\rm s}/\Omega,r_{\rm p}/R_1),\label{eq:Simple_edot}
	\end{align}
	where $\mu = M_1 M_2/M_{\rm t}$ is the reduced mass of the binary and,
	\begin{equation}
	t_d^{-1} \equiv \frac{T_0}{\mu a^2} k_2 \tau = \frac{2}{21} \left(\frac{M_2}{M_1}\right)\left(\frac{M_1+M_2}{M_1}\right)\left(\frac{R_1}{a}\right)^8 \gamma_{\rm eff}. \label{eq:td_def}
	\end{equation}
	The dimensionless functions $F_E$, $F_T$, and $F_{\rm ecc}$ are provided in equations (51-53) of VL20. Again, we have used the frequency independent $\gamma_{\rm f}$ as the f-mode viscous damping rate. In the weak friction limit, $F_E$, $F_T$, and $F_{\rm ecc}$ can be simplified to equations (46-48) of VL20.
	
% 	We have combined $F_E$ and $F_T$ to characterize the eccentricity evolution via,
% 	\begin{equation}
% 	F_{\rm ecc} = \frac{1}{9} \frac{(1-e^2)}{e^2}\left[\frac{F_E}{(1-e^2)}-F_T\right].
% 	\end{equation}
	We assume isotropic wind mass loss, such that the wind-driven secular time evolution of the semi-major axis is,
	\begin{equation}
	\left.\frac{\dot{a}}{a}\right|_{\rm Wind} = -\frac{\dot{M}_1}{M_t},\label{eq:adot_wind}
	\end{equation}
	and the eccentricity is unchanged.
	We define the circularization time of the binary as,
	\begin{equation}
	t_{\rm circ} \equiv \left|\frac{e}{\dot{e}}\right|. \label{eq:deftcirc}
	\end{equation} 
		
	As the binary circularizes, the tidal torque will cause the star to spin up to the pseudosynchronous rotation rate, where the star experiences no net torque. In the weak friction approximation, the pseudosynchronous rotation rate is given by, 
	\begin{equation}
	\Omega_{\rm ps} = \frac{f_2}{f_5(1-e^2)^{3/2}}\Omega, \label{eq:Omegaps}
	\end{equation}
	For a highly eccentric orbit, the true pseudosynchronous rotation rate, where $\dot{\Omega}_{\rm s}$ is zero, can occur at slightly faster rotation rates than given by equation~(\ref{eq:Omegaps}) (VL20). In a circular orbit, $f_2$ and $f_5$ are 1, and the synchronous rotation rate is the orbital frequency. 
	
	When the binary is sufficiently close, mass transfer will become important, and equations~(\ref{eq:Simple_adot}) - (\ref{eq:Simple_edot}) will no longer capture the orbital evolution. The onset of mass transfer occurs when the primary is Roche-lobe filling. A precise calculation of the Roche radius depends on both the stellar spin and orbital eccentricity \citep[e.g.][]{Sepinsky07}. For simplicity, we use a common approximation adapted from  \citet{Eggleton83},
	\begin{equation}
	r_{\rm Roche} = r_{\rm p} \frac{0.49 q^{2/3}}{0.6q^{2/3} + \ln(1+q^{1/3})},
	\end{equation}
	where $r_{\rm p}$ is the pericentre distance and $q = M_1/M_2$ is the mass ratio. For a binary with a companion of mass $M_2 = 1.4 M_\odot$, the primary fills its Roche-lobe when $r_{\rm p} \approx 1.8 R_1$ for $M_1 = 10 M_\odot$ and when $r_{\rm p} \approx 1.7 R_1 $ for $M_1 = 15 M_\odot$.
	
\section{Results}\label{sec:Results}
\subsection{Tidal Dissipation Rate as a Function of Stellar Evolution}\label{sec:circTime}
	We have used the MESA-generated stellar models introduced in Section~\ref{sec:Methods} to understand the tidal dissipation timescale of a star after it leaves the main sequence. 
	
\begin{figure*}
	\begin{center}
		\includegraphics[width = 3.2in]{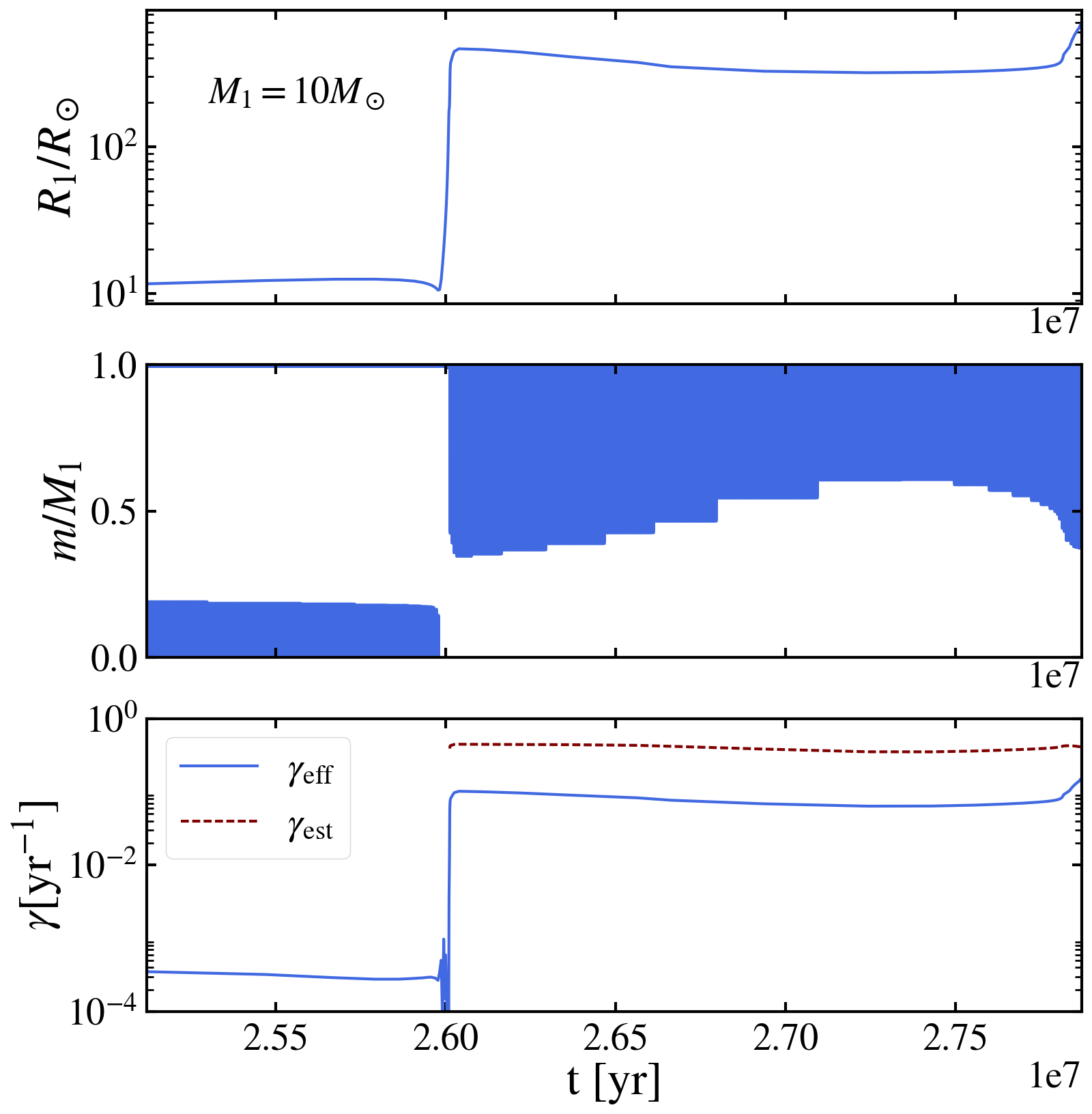}
		\includegraphics[width = 3.2in]{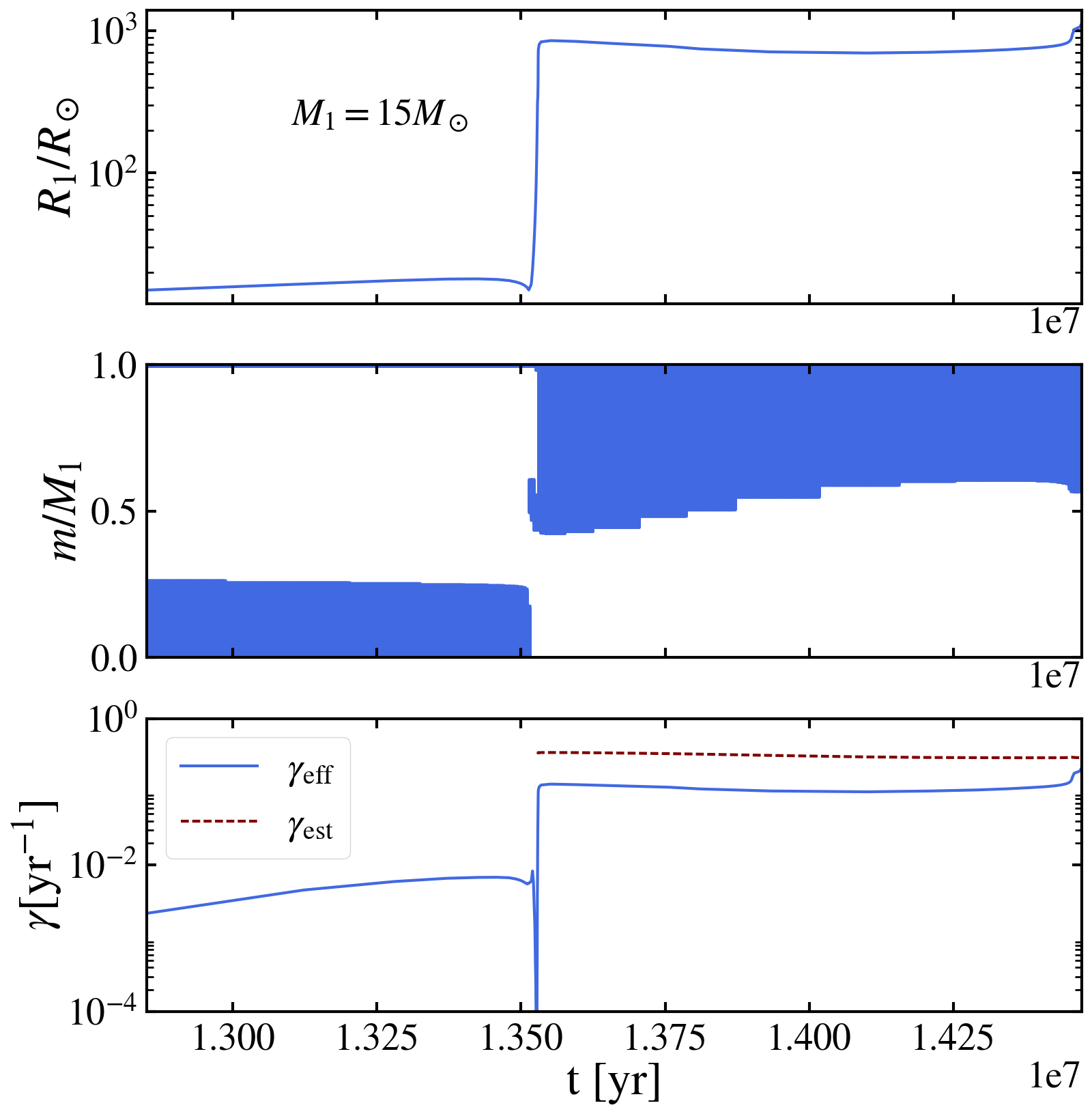}
		\caption{The evolution of the stellar structure and the effective tidal damping rate (from viscous dissipation in the convective envelope) for a $M_1 = 10M_\odot$ (left) and $M_1 = 15M_\odot$ (right) MESA-generated stellar model.The top panels show the evolution of the radius. The middle panels illustrate the convective regions in the stars. The bottom panels show the tidal damping rates for the stellar models calculated with equation~(\ref{eq:gammaEff}), together with equations~(\ref{eq:defk2}) and (\ref{eq:deftau}) and compared with the estimate from equation~(\ref{eq:gamma_est}).}
		\label{fig:StellarEvolution}
	\end{center} 
\end{figure*}

	Figure~\ref{fig:StellarEvolution} displays changes in the stellar radius and structure of the $10~M_\odot$ and $15~M_\odot$ stellar models as they evolve from core hydrogen burning to carbon depletion (at which point the collapse of the core is imminent). The top panel shows that the bulk of the radius evolution occurs in two bursts. On the giant branch, the radius expands by a factor of $\sim 40$ for the $10~M_\odot$ model and $\sim 60$ for the $15~M_\odot$ model. As the radius expands, the structure of the star changes significantly. The middle panels of Fig.~\ref{fig:StellarEvolution} show that the stars develop deep convective envelopes during the first episode of radius expansion. Once the envelope has developed, turbulent viscosity in the envelope is the most efficient mechanism for tidal dissipation in the star. The bottom panels of Fig.~\ref{fig:StellarEvolution} display the damping rate from viscous dissipation, calculated via equation~(\ref{eq:gammaEff}) together with equations~(\ref{eq:defk2})-(\ref{eq:deftau}) and $\gamma_{\rm f}$ from equation (19) of VL20. This calculation assumes a significant convection zone within the star, and likely does not capture the tidal dissipation rate of the star before the envelope begins to develop. We can compare the more precise calculation of $\gamma_{\rm eff}$ with the convenient estimate from \cite{Zahn75,Phinney92,Verbunt95} provided in equation~(\ref{eq:gamma_est}), and find that, while $\gamma_{\rm eff}$ is smaller than $\gamma_{\rm est}$, they agree within a factor of a few.
	
\begin{figure*}
	\begin{center}
		\includegraphics[width = 3.2in]{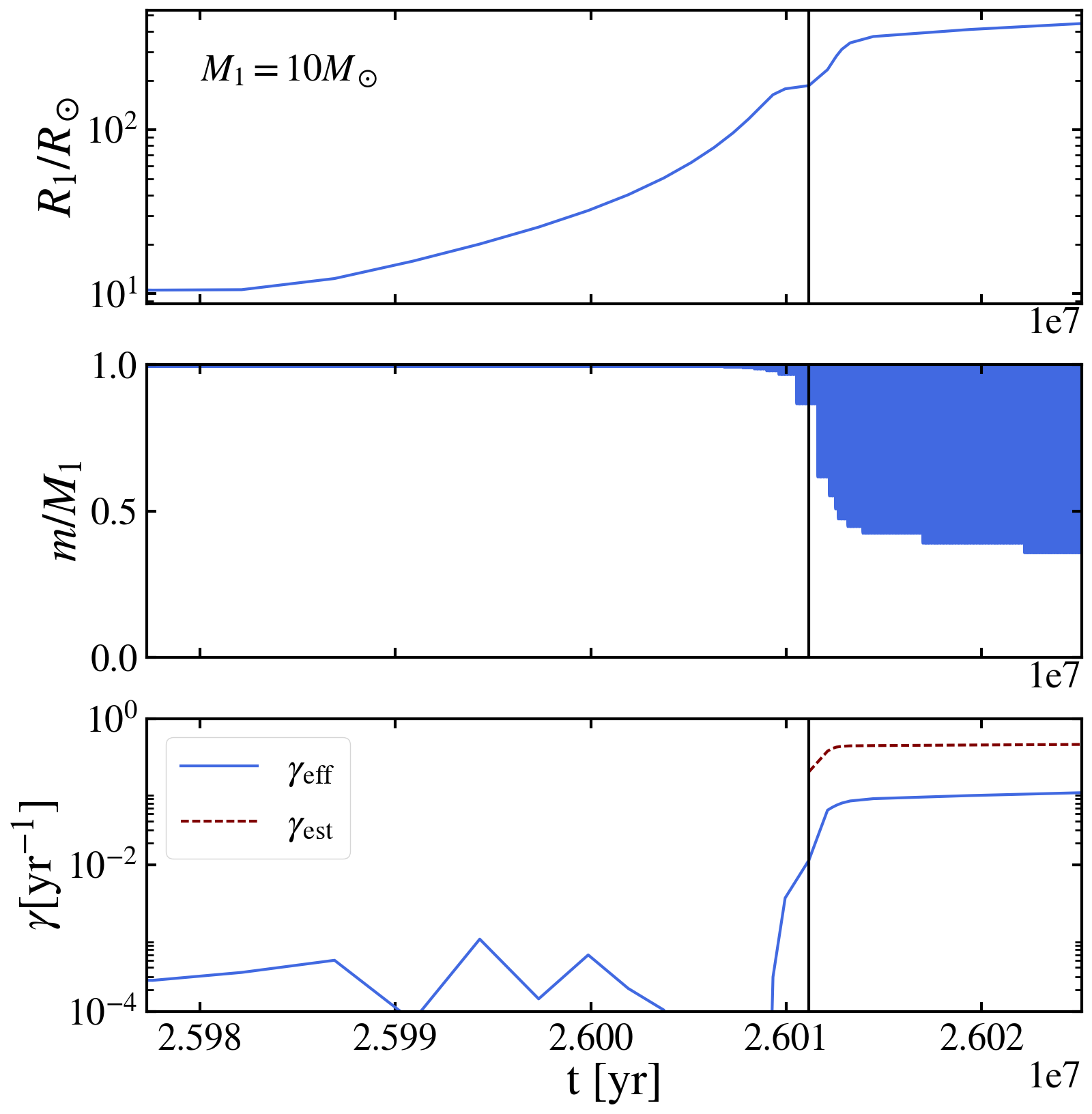}
		\includegraphics[width = 3.2in]{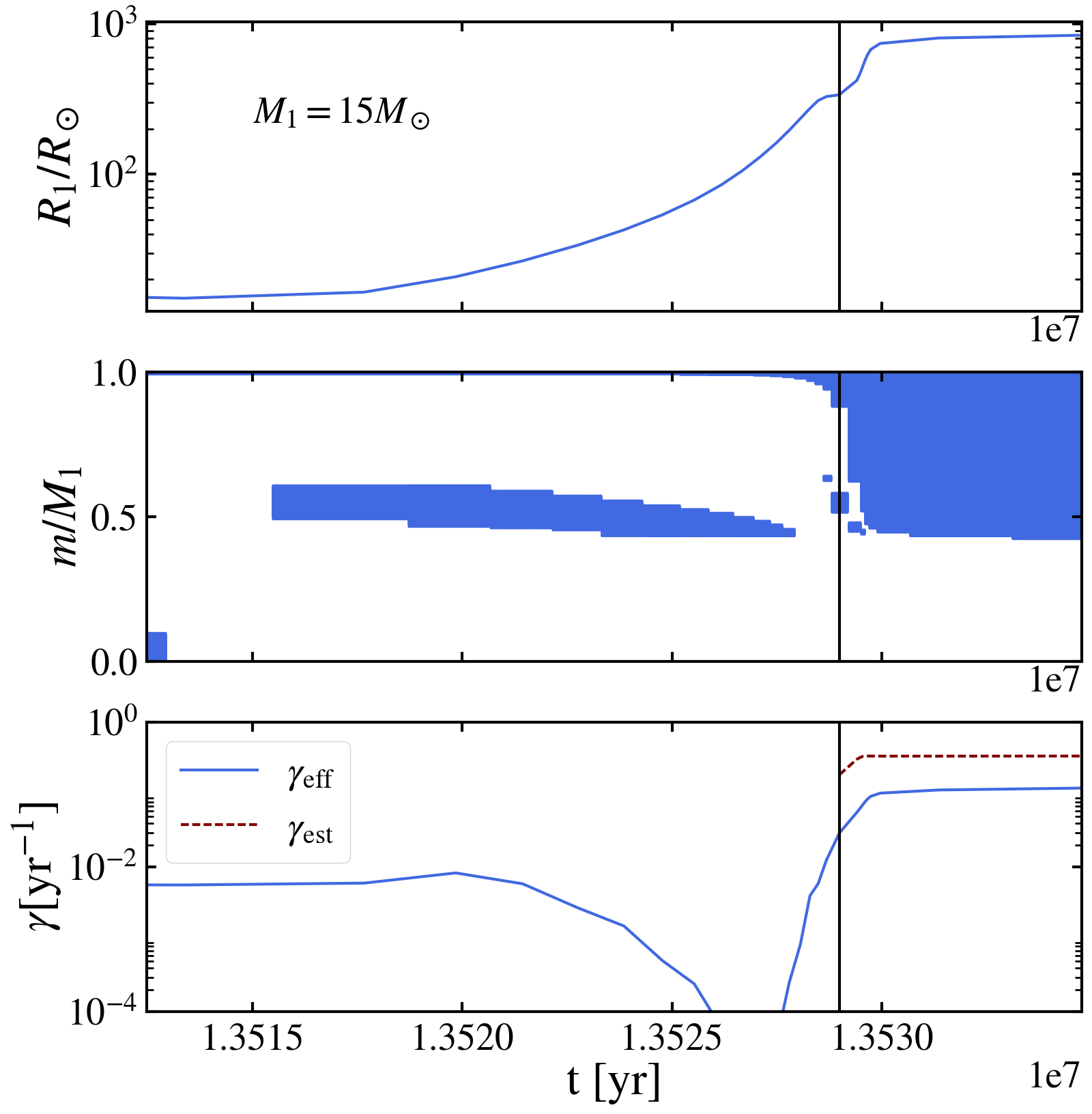}
		\caption{The same as Fig.~\ref{fig:StellarEvolution} but zoomed-in on the first episode of radius expansion. The vertical black lines mark the development of a convective envelope with $M_{\rm env}>0.1M_1$. The viscous dissipation rates shown in the bottom panels likely do not capture tidal dissipation to the left of the black line before the development of the envelope.}
		\label{fig:Zoom}
	\end{center}
\end{figure*}

	For many binaries, the timing of when deep convection of the outer envelope develops in the primary will determine whether tides can play an important role in circularizing the orbit before the onset of RLO. Before the convective envelope forms, tidal dissipation cannot circularize the orbit efficiently because the timescale for viscous dissipation is longer than the stellar evolution timescale. Figure~\ref{fig:Zoom} provides a zoom-in of the first episode of radius expansion in Fig.~\ref{fig:StellarEvolution}. The vertical black lines indicate the development of a significant convective envelope with $M_{\rm env } > 0.1 M_1$. For the $10~M_\odot$ ($15~M_\odot$) stellar model, this criterion is met when the radius has expanded by a factor of $\sim 18$ ($\sim26$). Following helium ignition in the core, the stellar radius will further expand to $\sim 43$ ($\sim 66$) times its value at the end of the main sequence over the course of about $2\times 10^5$~yrs ($2\times 10^4$~yrs). 
	
\subsection{Tidal Circularization Timescale and the Binary Orbit}
\begin{figure}
	\begin{center}
		\includegraphics[width = \columnwidth]{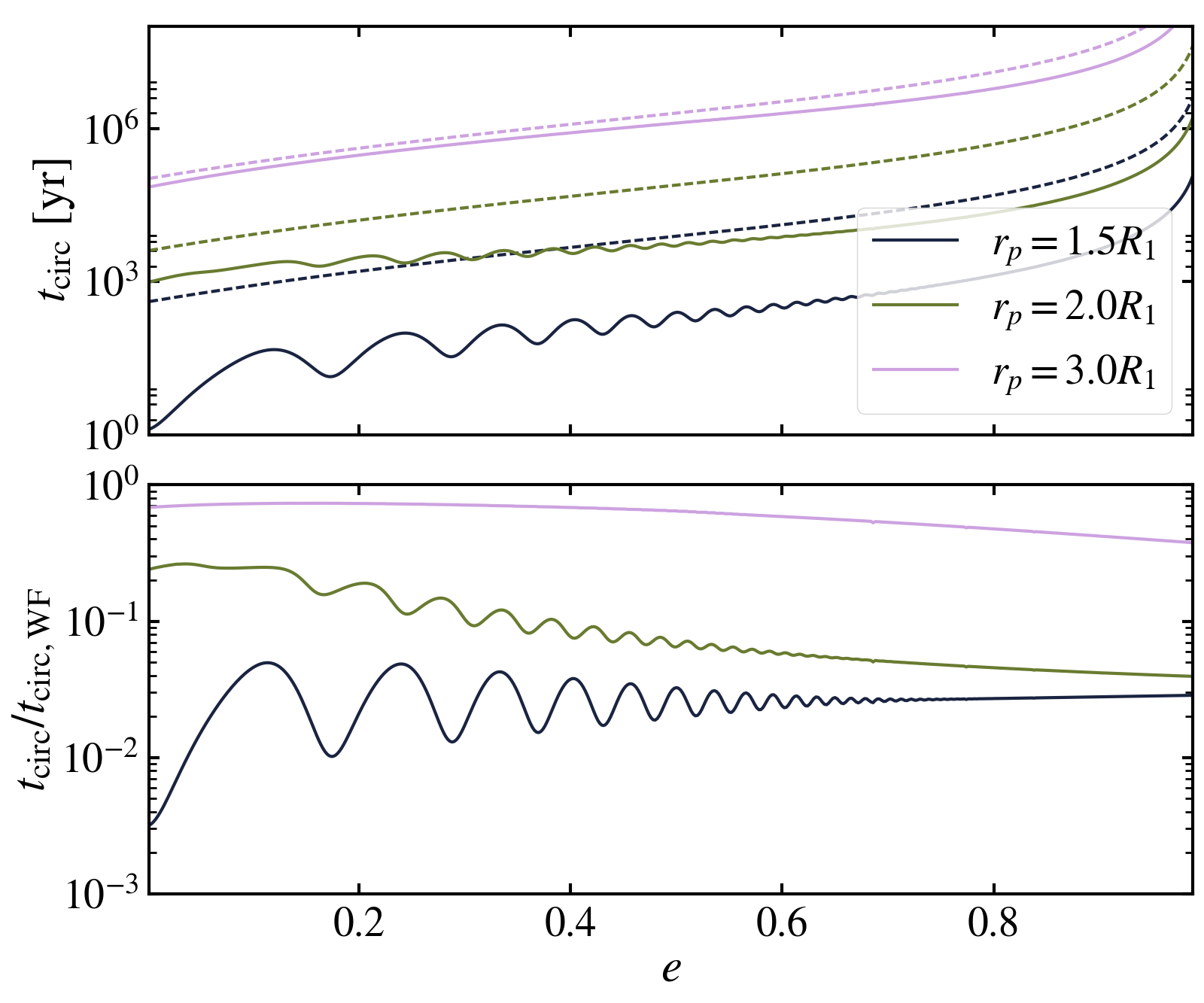}
		\caption{Circularization timescale ($|e/\dot{e}|$) from equation~(\ref{eq:Simple_edot}) at a variety of pericentre distances for a binary with a $1.4 M_\odot$ companion as a function of eccentricity. The timescales were calculated with $\gamma_{\rm eff}$=0.4 yr$^{-1}$, $\omega_{\rm f}= 1.75(GM_1/R_1^3)^{1/2}$, and $\gamma_{\rm f} =0.04(GM_1/R_1^3)^{1/2}$. The bottom panel shows the ratio of the timescale to the weak friction timescale, given by equation~(\ref{eq:Simple_edot}) with $F_{\rm ecc}$ from equation(48) of VL20. As $r_{\rm p}$ increases, the ratio asymptotes to 1.}
	\label{fig:tcirc}
	\end{center}
\end{figure}

		The tidal circularization timescale depends not only on the stellar structure, but also the binary orbital parameters and the spin rate of the primary. Figure~\ref{fig:tcirc} shows the tidal circularization time $t_{\rm circ}$ (see equations~\ref{eq:Simple_edot} and \ref{eq:deftcirc}) as a function of the orbital eccentricity for a variety of pericentre values. The timescale is calculated for a non-rotating $10~M_\odot$ primary that has already ascended the giant branch and developed deep convection in a binary with a $1.4~M_\odot$ companion. We have used $\gamma_{\rm eff} = 0.4~{\rm yr}^{-1}$, and set the f-mode frequency and damping rate to $\omega_{\rm f} = 1.75 (GM_1/R_1^3)^{1/2}$ and $\gamma_{\rm f} = 0.04 (GM_1/R_1^3)^{1/2}$. 
		
		The circularization time depends very strongly on $r_{\rm p}/R$. For a given $r_{\rm p}$, this timescale is shortest for small orbital eccentricities. In Fig.~\ref{fig:tcirc}, the circularization timescales for the two smallest pericentre distances, $r_{\rm p}/R_1 = 1.5$ and $r_{\rm p}/R_1 = 2.0$, have a complicated dependence on the orbital eccentricity. This arises due to enhanced tidal dissipation at resonances between the orbital period and the f-mode frequency in the inertial frame. 
		
		Figure~\ref{fig:tcirc} also displays the standard weak friction calculation of the circularization timescale $t_{\rm circ,WF}$ in the top panel (dashed lines). The bottom panel shows the ratio of $t_{\rm circ}$ to $t_{\rm circ,WF}$. The weak friction timescale is given by equation~(\ref{eq:Simple_edot}), but calculated with the dimensionless functions $F_{T}$ and $F_{E}$ from equations~(46) and (47) of VL20. The weak friction approximation is valid when the dominant tidal forcing frequency, of order the pericentre frequency $\Omega_p = \Omega (1+e)^{1/2}/(1-e)^{3/2}$, is much slower than the f-mode frequency, which is of order $(GM_1/R_1^3)^{1/2}$. At large $r_{\rm p}/R_1$ and small eccentricity, where $\Omega_p \ll  (GM_1/R_1^3)^{1/2}$, the two timescales $t_{\rm circ}$ and $t_{\rm circ,WF}$ are in good agreement. However, for larger $e$ and smaller $r_{\rm p}$ the weak friction calculation overestimates the circularization timescale by a few orders of magnitude. 
		
\begin{figure*}
	\begin{center}
		\includegraphics[width = 4.5in]{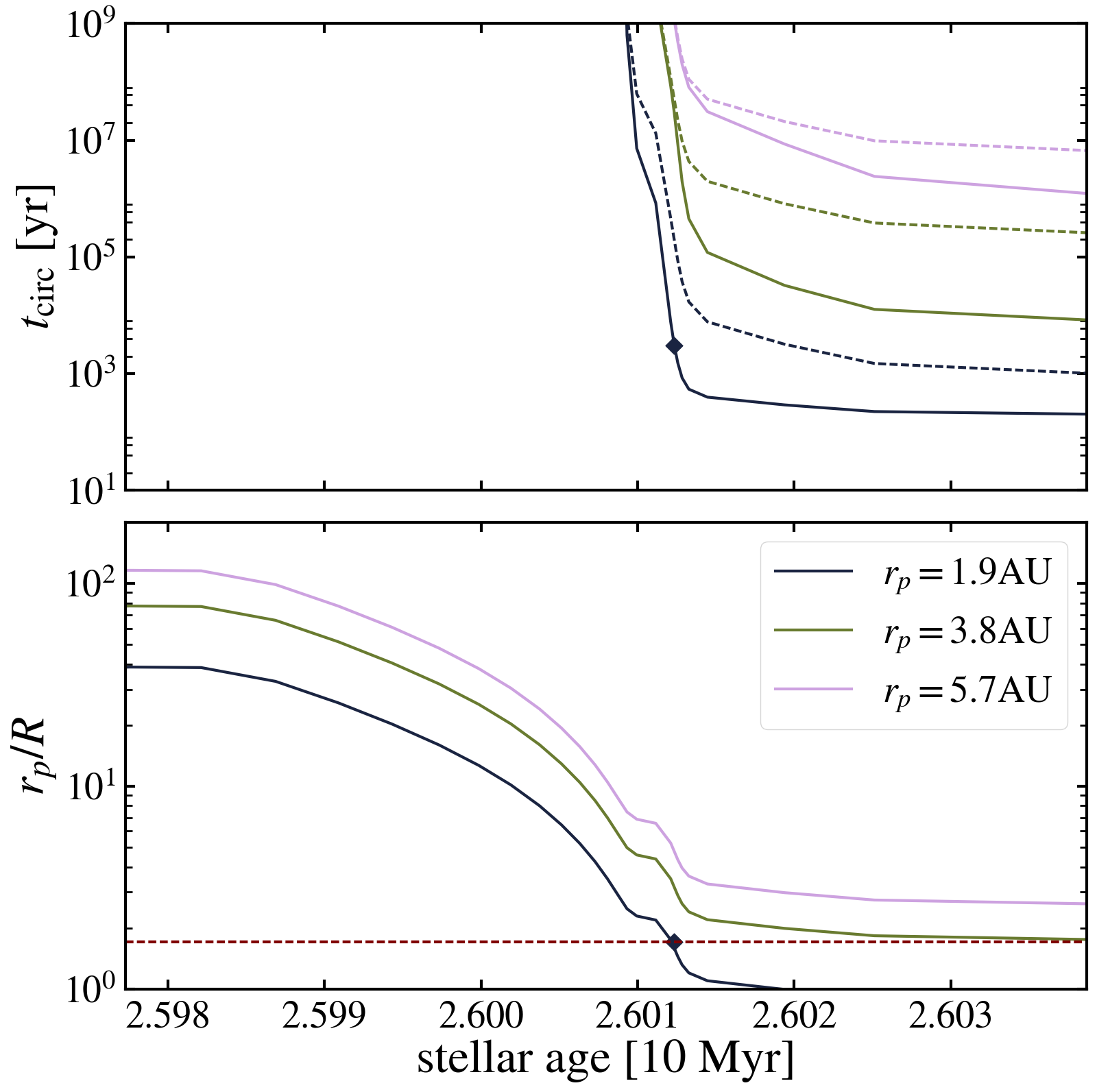}
		\caption{Top panel: The circularization timescale ($\tau_{\rm circ} = |e/\dot{e}|$) for a binary with   $M=10 M_\odot$ primary, a $1.4 M_\odot$ companion, and initial eccentricity $e_0=0.8$. Each solid line shows the timescale calculated via equations~(\ref{eq:Simple_edot}) and (\ref{eq:deftcirc}), and dashed lines of the same color show the weak friction result for comparison. The black dotted line indicates when the convective envelope develops ($M_{\rm env} > 0.1 M_1$).
		This calculation considers only changes in the stellar structure and radius. Bottom panel: The pericentre distance in units of the evolving stellar radius. The horizontal red dashed line is the condition for RLO.}
		\label{fig:tcircEvolutionEccp9}
	\end{center}
\end{figure*}

		As the radius of the primary expands, $r_{\rm p}/R_1$ rapidly decreases. Figure~\ref{fig:tcircEvolutionEccp9} shows $t_{\rm circ}$ for the $10~M_\odot$ stellar model assuming a fixed orbit with $e=0.8$ and a given pericentre distances ($r_{\rm p} = 1.9, 3.8,$ or $5.7$~au), and a non-rotating but evolving primary. The convective envelope does not develop until a little after $2.6\times 10^7$~yrs (indicated with a dotted black line in Fig.~\ref{fig:tcircEvolutionEccp9}). The dashed black red in the bottom panel is the Roche limit. The small black diamond indicates the onset of RLO for the $r_{\rm p} = 1.9$~au binary. In this system, the star reaches $r_{\rm Roche}$ within $\sim2 \times 10^3$ years of developing a convective envelope. This timescale is faster than $t_{\rm circ}$ for the binary (before crossing the red dashed line). Although the tidal circularization time is shortest for small $r_{\rm p}/R$, tight binaries also have the shortest timescales for reaching RLO. Systems where $r_{\rm p}$ is too small will not have time to tidally circularize before the primary grows to $r_{\rm Roche}$.
		
\subsection{Outcomes of Coupled Stellar and Orbital Evolution} \label{sec:Outcome}
\begin{figure*}
		\begin{center}
			\includegraphics[width = 4.5 in]{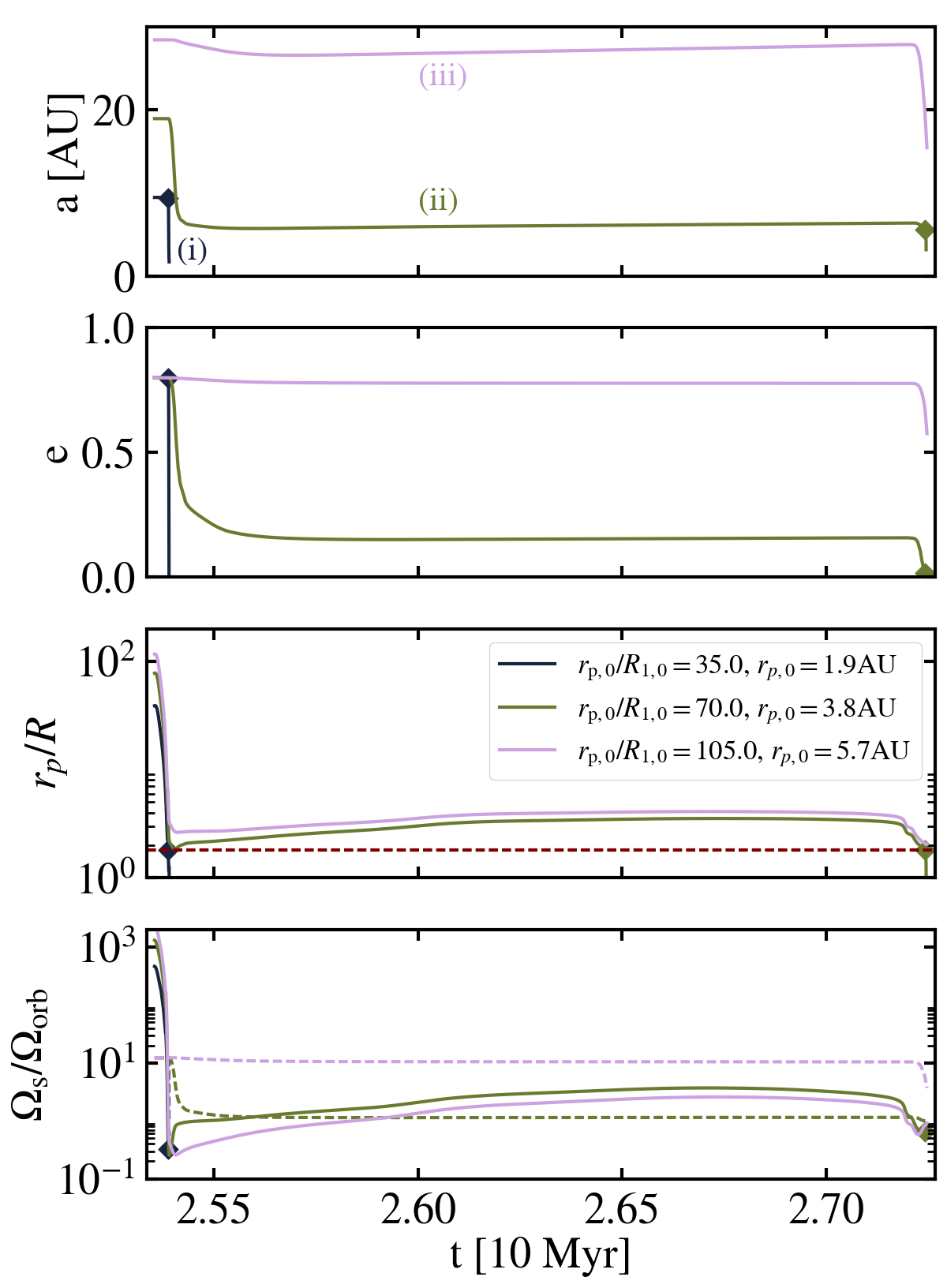}
			\caption{The orbital evolution of binary with a $10 M_\odot$ primary star, initial eccentricity $e_0 = 0.8$, and the same initial pericentre values $r_{\rm p,0}$ as in Fig.~\ref{fig:tcircEvolutionEccp9}. The cases are numbered according to the outcomes listed at the start of Section~\ref{sec:Outcome}. The top two panels are $a$ and $e$. The third panel is $r_{\rm p}/R_1$, with the Roche radius $r_{\rm p} = r_{\rm Roche}$ indicated by the red dashed line. The bottom panel shows the rotation period of the star (solid lines) and the weak friction pseudosynchronous rotation rate from equation~(\ref{eq:Omegaps}) (dashed lines). The diamonds in each panel indicate when a given binary reaches the criterion for RLO.}
			\label{fig:3Cases}
		\end{center}
	\end{figure*}

	Next, we present the results of coupled stellar and orbital evolution for binaries using the $10$ and $15~M_\odot$ stellar models and the method described in Section~\ref{sec:Methods}. We assume a compact companion with $M_2=1.4 M_\odot$. Figure~\ref{fig:3Cases} highlights three possible outcomes of coupled orbital and stellar evolution. These are:
	\begin{enumerate}[leftmargin=\parindent]
	    \item a binary that does not circularize before reaching the Roche-radius;
	    \item a binary that is tidally circularized before the primary evolves to $R_1 = r_{\rm Roche}$; and
	    \item a system that is too wide to undergo RLO.
	\end{enumerate}
	In each case, the initial orbital eccentricity is $e_0=0.8$ and the initial stellar rotation period is 1 day. This choice of rotation period is motivated by the observation that the average surface rotation rate of B stars is $\sim 25\%$ of breakup \citep{Abt02}. The three different values of the initial pericentre separation $r_{\rm p,0}$ are the same as those used in Fig.~\ref{fig:tcircEvolutionEccp9}. 
	
	For the smallest value of $r_{\rm p,0}$ the orbit does not circularize before the star grows to $R_1 = r_{\rm Roche}$.
	There is not enough time between the development of the convective envelope and reaching the Roche radius for tidal dissipation in the envelope to circularize the orbit. The black diamonds in Fig.~\ref{fig:3Cases} mark the binary parameters and stellar rotation rate when $R_1 = r_{\rm Roche}$. At this point the semi-major axis and eccentricity are essentially unchanged, while the rotation rate has slowed to conserve the spin angular momentum of the star as the radius expands.
	
	In the intermediate case where $e_0=0.8$ and $r_{\rm p,0} = 70 R_{1,0}$, the orbit circularizes significantly as the primary climbs the giant branch. The orbit continues circularizing when the primary ascends the asymptotic branch. In this case, the binary is nearly circular when $R_1 = r_{\rm Roche}$. The rotation period of the primary is slightly longer than the predicted pseudosynchronous rate from the weak friction theory. As the star expands, its rotation slows down. This effect is in competition with the tidal torque, which acts to spin up the convective envelope. The resulting stellar rotation rate at $R_1 = r_{\rm Roche}$ is $\sim 1.5$ times slower than the orbital frequency (which is the expected pseudosynchronous rotation rate for a nearly circular orbit).
	
	In the final case, displayed as a pink line in Fig.~\ref{fig:3Cases}, the initial pericentre distance is quite large. The binary cannot reach the criterion for RLO within the lifetime of the primary star. If we take the Roche radius as a hard limit for mass transfer, this binary is too widely separated to undergo a common envelope episode. 
	
	With an understanding of the main categories of outcomes, we can survey results from a range of orbital configurations. Figures~\ref{fig:FinalEcc_M10} and~\ref{fig:FinalEcc_M15} display the binary eccentricity at the onset of RLO, $e_{\rm Roche}$, for a variety of `initial' orbital parameters when the primary star evolves off of the main sequence. We have assumed that stellar rotation period at this point is 1 day, but found that altering the initial spin has a negligible effect on $e_{\rm Roche}$. For the $10~M_\odot$ stellar model, binaries with $r_{\rm p,0} \lesssim 3$~AU on the main sequence do not circularize before reaching the Roche radius. The same is true of binaries with $r_{\rm p,0} \lesssim 6$~AU for the $15~M_\odot$ stellar model. The orbital evolution of these binaries looks something like the black line from Fig.~\ref{fig:3Cases}. Though the timescale for tidal circularization can become very short as the stellar radius expands, it is still longer than the time between when the primary develops a convective envelope and when the radius grows to the Roche limit. In some cases, the star cannot even develop a convective envelope before mass transfer begins (systems below the white dashed line in Figs.~\ref{fig:FinalEcc_M10} and \ref{fig:FinalEcc_M15}). These binaries retain their initial eccentricity from before the primary leaves the main sequence. 
	
	In contrast, many binaries with an initial separation greater than $\sim4$~AU for a $10~M_\odot$ primary or $\sim7$~AU for $ M_1 = 15~M_\odot$ circularize completely before RLO begins. In these binaries, the timescale between the onset of deep convection in the primary and radius growth to $R_1 = r_{\rm Roche}$ is longer than the tidal circularization timescale. The green line in Fig.~\ref{fig:3Cases} is an example of a system in this category.
	
	Some binaries with large pericentre  distances and high eccentricities will never undergo RLO. In Figs.~\ref{fig:FinalEcc_M10} and \ref{fig:FinalEcc_M15}, these systems lie above the solid black lines. They have the longest timescales for tidal circularization. 
	
	Figures~\ref{fig:FinalEcc_M10} and \ref{fig:FinalEcc_M15} also display the rotation rate of the primary at the Roche radius as a fraction of the pseudosynchronous rotation rate (see equation~\ref{eq:Omegaps}). Before the primary develops a convective envelope, tides are inefficient at transferring angular momentum between the orbit and the star. The spin angular momentum of the star is effectively conserved, so the stellar rotation rate slows as the radius expands. For the $10~M_\odot$ ($15~M_\odot$) stellar model, in a binary with $r_{\rm p,0} \lesssim 3$~AU ($6$~AU) the stellar rotation rate at the Roche radius is given by $\Omega_{\rm s} = (k_0 M_{1,0} R_{1,0}^2) \Omega_{\rm s,0}/(k M_1 R_1^2)$, where the $0$ subscript indicates the value when the star leaves the main sequence. When the star develops a convective envelope, its structure changes significantly as does the moment of inertia constant $k$. This change accounts for the sharp transition in the stellar rotation rate across the white dashed line in the right panels of Figs.~\ref{fig:FinalEcc_M10} and \ref{fig:FinalEcc_M15}. Both Figs.~\ref{fig:FinalEcc_M10} and \ref{fig:FinalEcc_M15} show a light green curve in $e_0$ and $r_{\rm p,0}$ below the black line that separates wide binaries (though this feature is far more obvious in Fig.~\ref{fig:FinalEcc_M10}). Systems above this ridge are nearly pseudosynchronous at the onset of RLO. Below the ridge, the primary star grows to the Roche radius while on the giant branch. Above, the pimary reaches $R_1 = r_{\rm Roche}$ on the asymptotic branch.
	
	When $r_{\rm p,0}$ is slightly larger, tidal dissipation can alter the orbit before $R_{1} = r_{\rm Roche}$, and the tidal torque spins up the star to a rotation rate of order the orbital frequency. Note that this effect acts in opposition to the radius expansion, which decreases the stellar spin rate. 

\subsection{Tidal Circularization and the Eccentricity Distribution of pre-Common Envelope Stellar Binaries}
	We use the results from the Section~\ref{sec:Outcome} to understand how tidal dissipation should affect the distribution of binary eccentricities and stellar rotation rates at the onset of RLO.  
	
\begin{figure*}
			\begin{center}
				\includegraphics[width = 3.25 in]{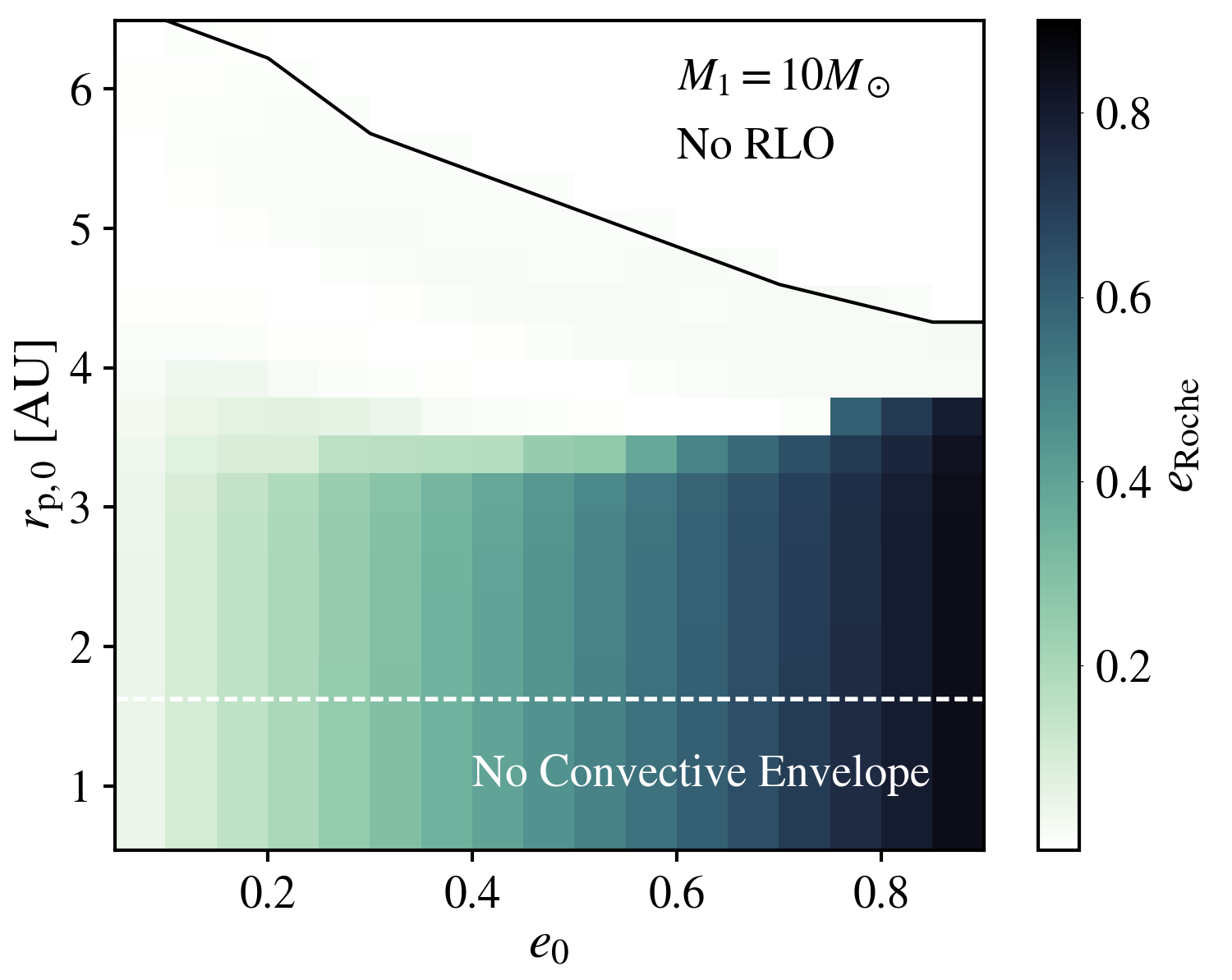}
				\includegraphics[width = 3.25 in]{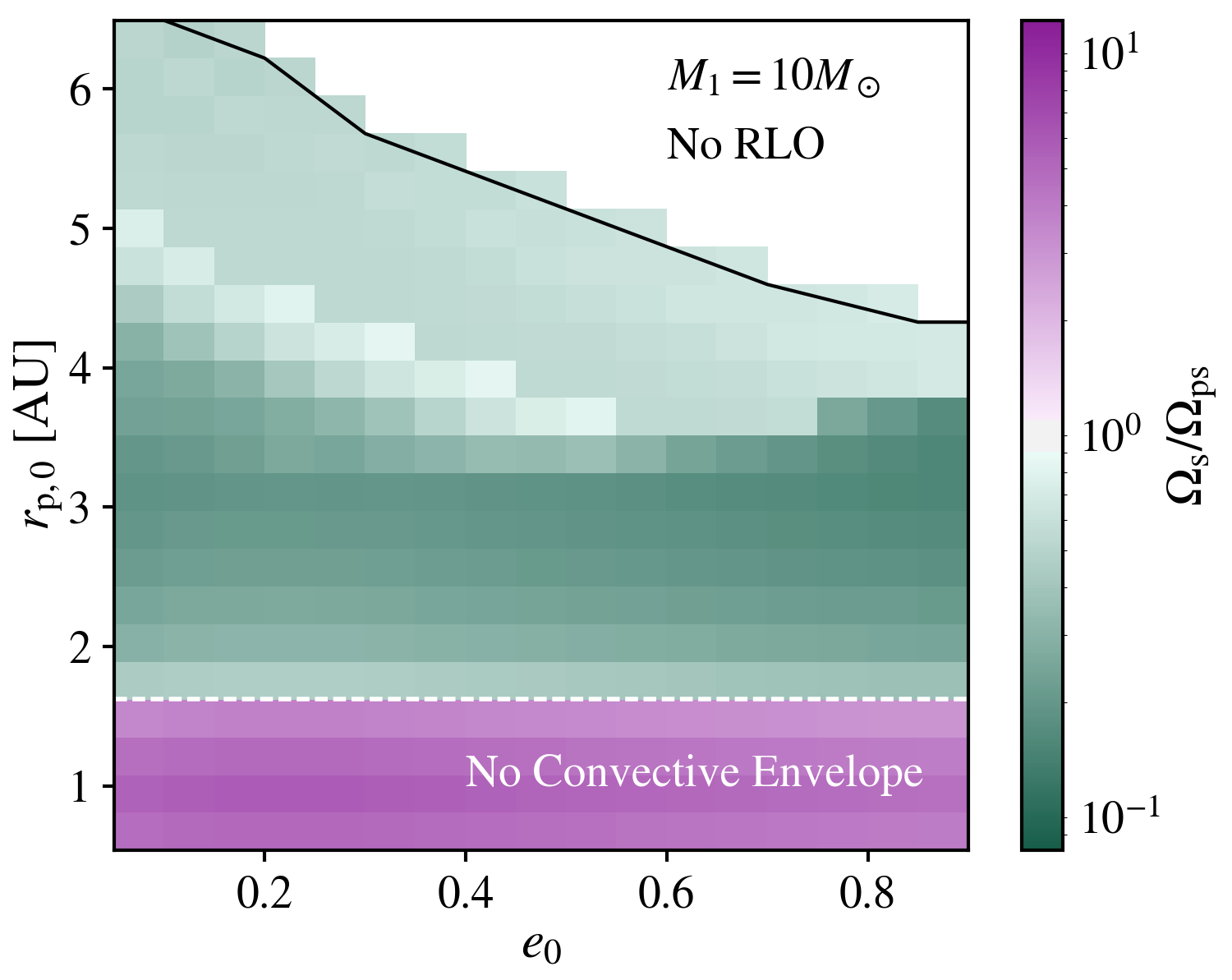}
				\caption{ Left: Eccentricity at the Roche radius for a $10 M_\odot$ primary star with a $1.4~M_\odot$ companion ( $r_{\rm p} = 1.8 R_1$). Right: The ratio of the stellar rotation rate $\Omega_{\rm s}$ to the weak friction pseudosynchronous rate $\Omega_{\rm ps}$ (see equation~\ref{eq:Omegaps}). Both panels are shown in the parameter space of the initial (main-sequence) pericentre distance and eccentricity. Systems above the solid black line are too wide to merge within the lifetime of the primary star. Below the dashed white line, $r_{\rm p,0}$ is too small for the star to develop a convective envelope before $r_{\rm p} = r_{\rm Roche}$.}
				\label{fig:FinalEcc_M10}
			\end{center}
		\end{figure*}

\begin{figure*}
			\begin{center}
				\includegraphics[width = 3.25 in]{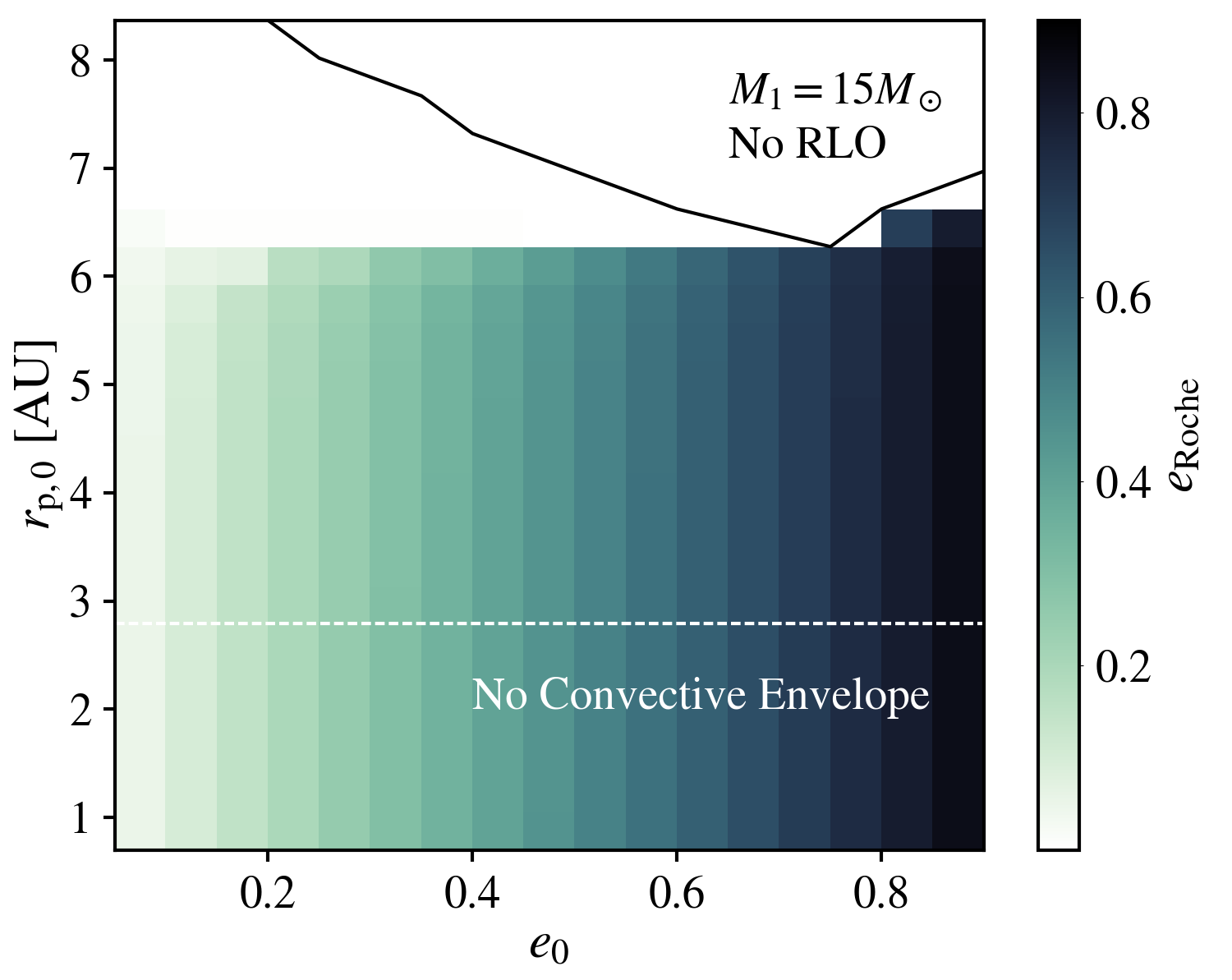}
				\includegraphics[width = 3.25 in]{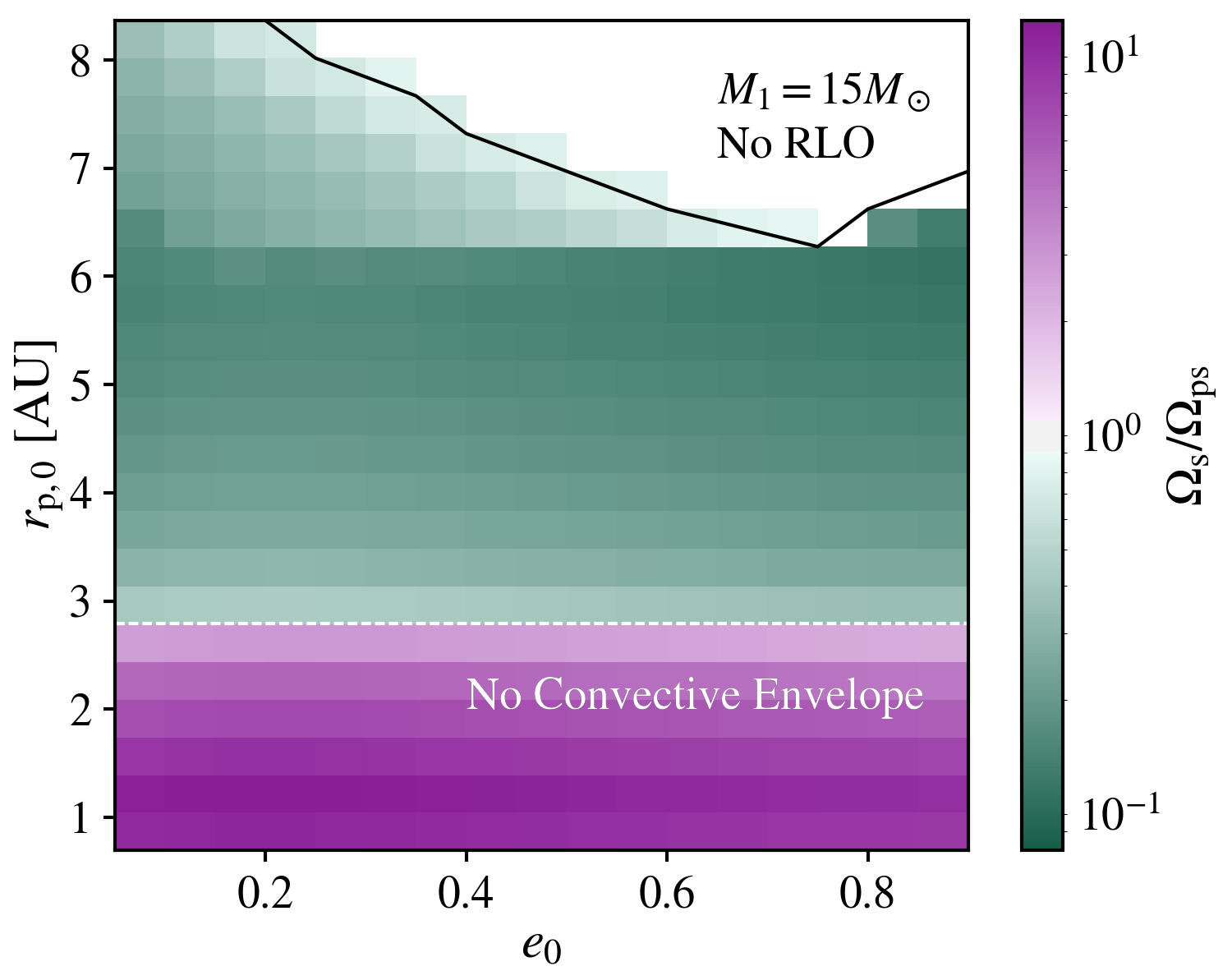}
				\caption{Same as in Fig.~\ref{fig:FinalEcc_M10} but for the $15 M_\odot$ stellar model.}
				\label{fig:FinalEcc_M15}
			\end{center}
		\end{figure*}

\begin{figure*}
			\begin{center}
				\includegraphics[width = 3.25 in]{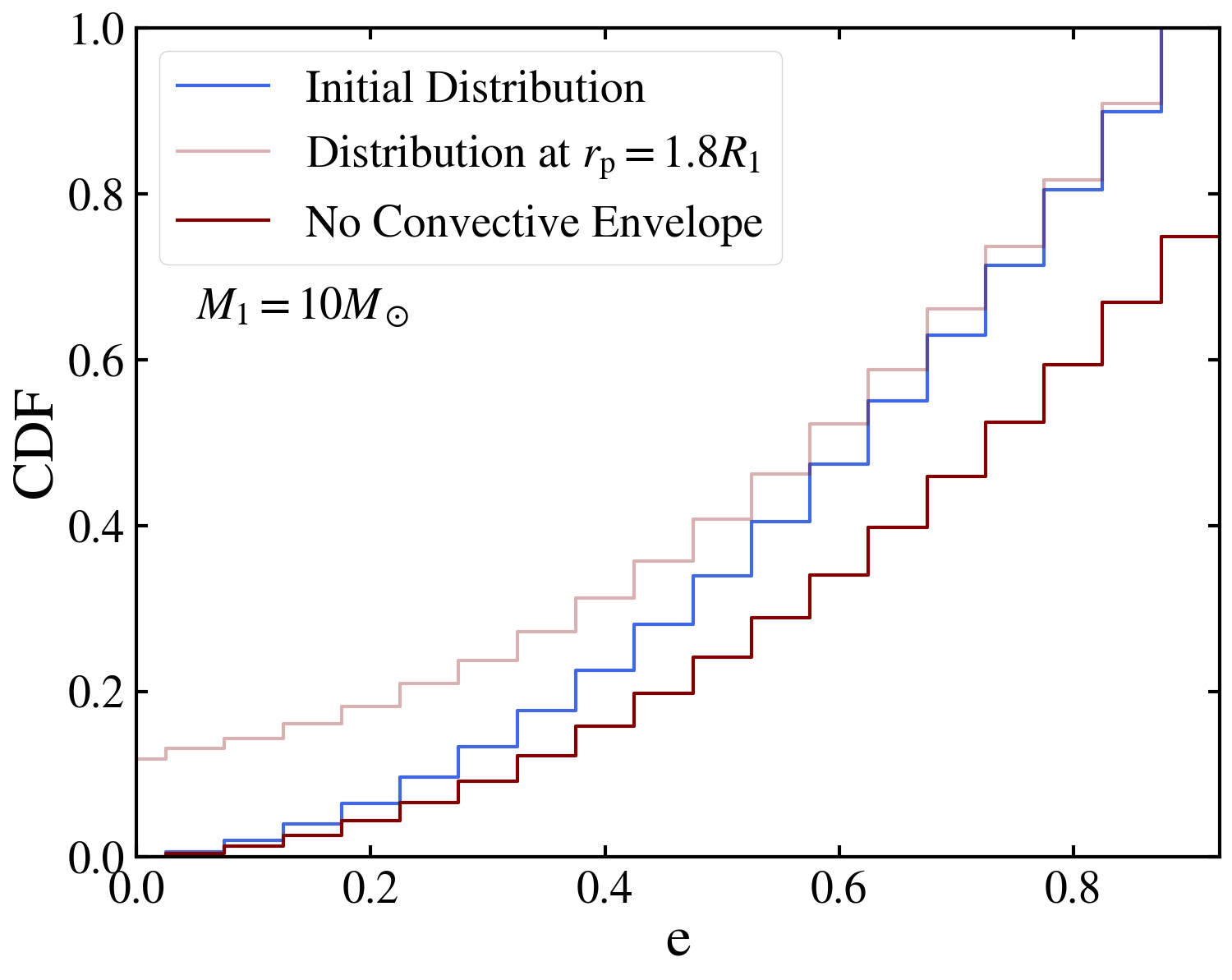}
				\includegraphics[width = 3.25 in]{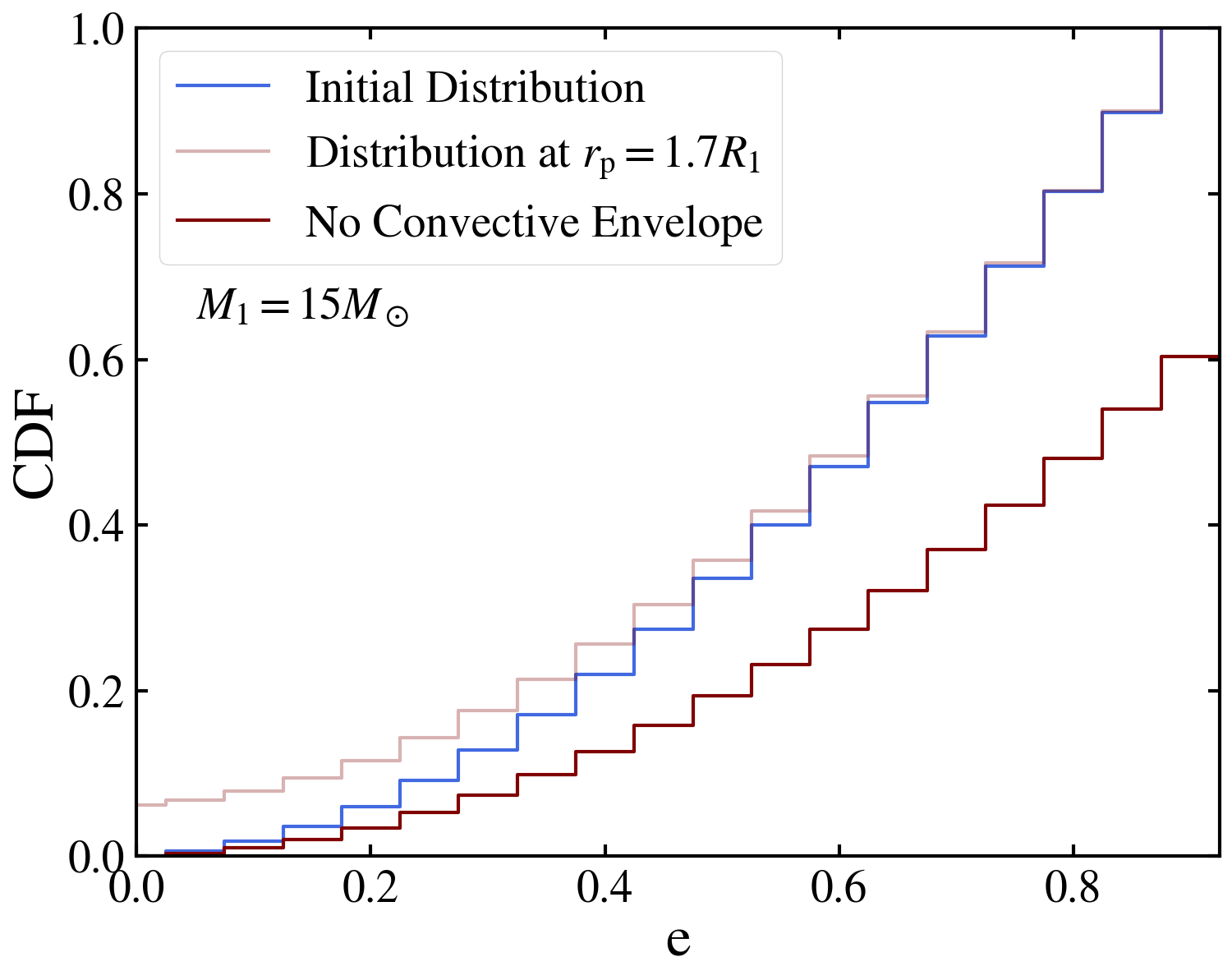}
				\caption{The cumulative distribution function of the binary eccentricities at the Roche radius for the $10 M_\odot$ stellar model (left) and $15 M_\odot$ stellar model (right) given a thermal initial eccentricity distribution. The light pink line is the eccentricity distribution of binaries for which $r_{\rm Roche}$ is smaller than the stellar radius at which the star develops a convective envelope ($M_{\rm env} > 0.1 M_1$).}
				\label{fig:HistogramRoche}
			\end{center}
		\end{figure*}

	For simplicity, we begin with a thermal distribution in eccentricity \citep{Jeans1919} and a log-uniform distribution in semi-major axis \citep{Opik2024} (i.e. $e^2$ is drawn uniformly from $[0,1]$ and $\log_{10}(a/{\rm AU})$ from $[-1,3]$). We then reject any set of initial conditions with an eccentricity and pericentre outside of the range spanned by the grid of results in Figs.~\ref{fig:FinalEcc_M10} and \ref{fig:FinalEcc_M15}. This provides an initial distribution of $r_{\rm p,0}$ and $e$ within the ranges $7.5 R_{1,0} < r_{\rm p,0} <122.5 R_{1,0}$ and $0.025 < e < 0.925$. The initial eccentricity distribution is shown as the blue line in Fig.~\ref{fig:HistogramRoche}. We then group the initial orbital parameters into bins of width $0.05$ in eccentricity and $5 R_{1,0}$ in pericentre distance that correspond to the grid in Figs.~\ref{fig:FinalEcc_M10} and \ref{fig:FinalEcc_M15}. We use the corresponding value of $e_{\rm Roche}$ for each bin in Figs.~\ref{fig:FinalEcc_M10} or \ref{fig:FinalEcc_M15} to create a distribution of eccentricities at the Roche radius. Binaries that never reach RLO are removed from both the initial eccentricity distribution and the Roche radius distribution. The resulting cumulative distribution functions for the two stellar models are shown as the light pink lines in Fig.~\ref{fig:HistogramRoche}. The dark red line shows the contribution to the eccentricity distribution from binaries where $r_{\rm p,0}$ is too small for the primary to develop a convective envelope before $R_1 = r_{\rm Roche}$.
	
	For both the $10~M_\odot$ and $15~M_\odot$ stellar models, the distribution of $e_{\rm Roche}$ is shifted toward lower eccentricities. For our choice of initial distribution of binary properties, $e_{\rm Roche} <0.1$ for $\sim 13 \%$ of systems that reach RLO for the $10 M_\odot$ star ($\sim 4 \%$ for the $15 M_\odot$ star). However, higher values of $e_{\rm Roche}$ still contribute significantly to the distribution. This is consistent with the fact that many binaries will not have time to circularize before reaching the RLO criterion. 
	
\begin{figure*}
			\begin{center}
				\includegraphics[width = 3.25 in]{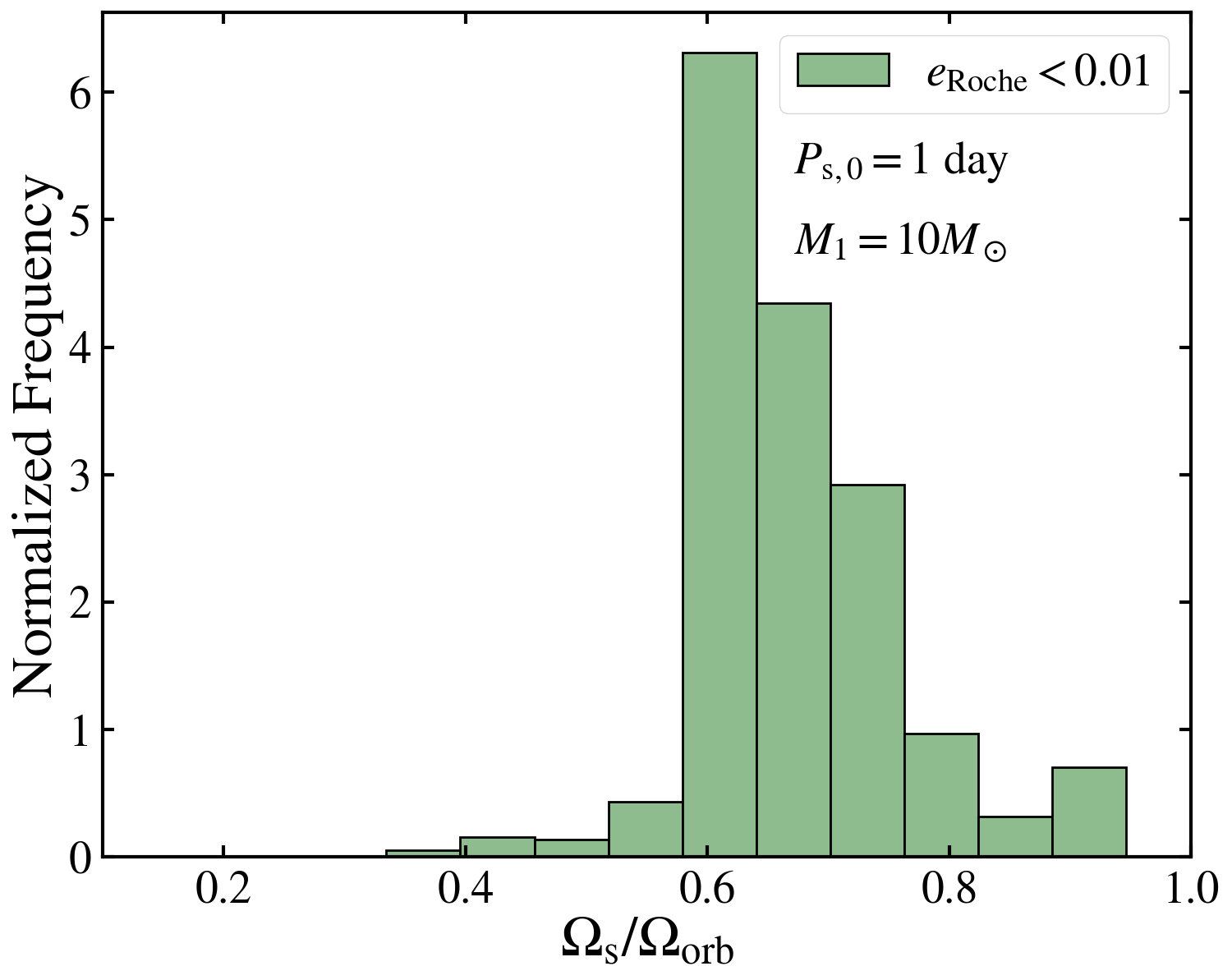}
				\includegraphics[width = 3.25 in]{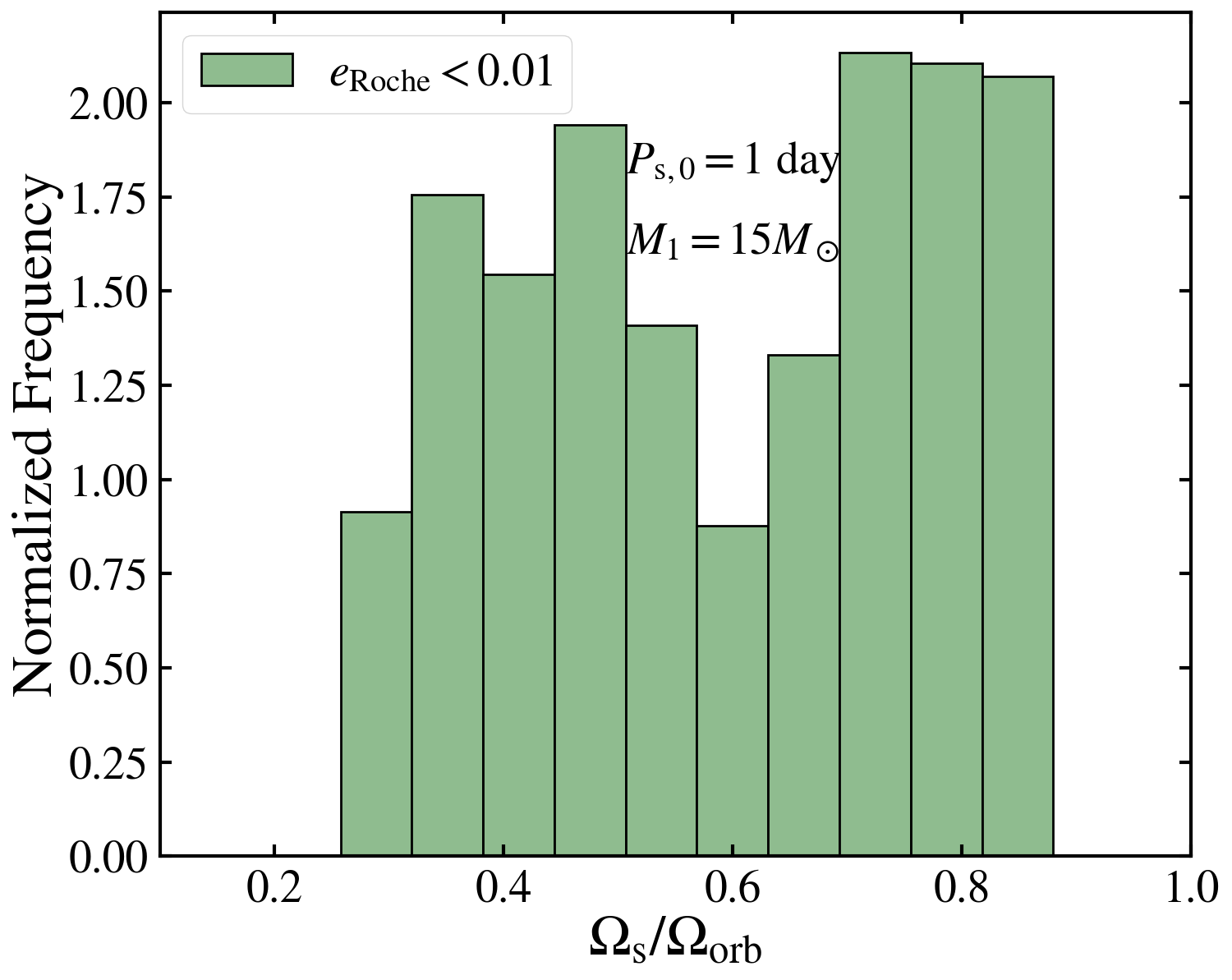}
				\includegraphics[width = 3.25 in]{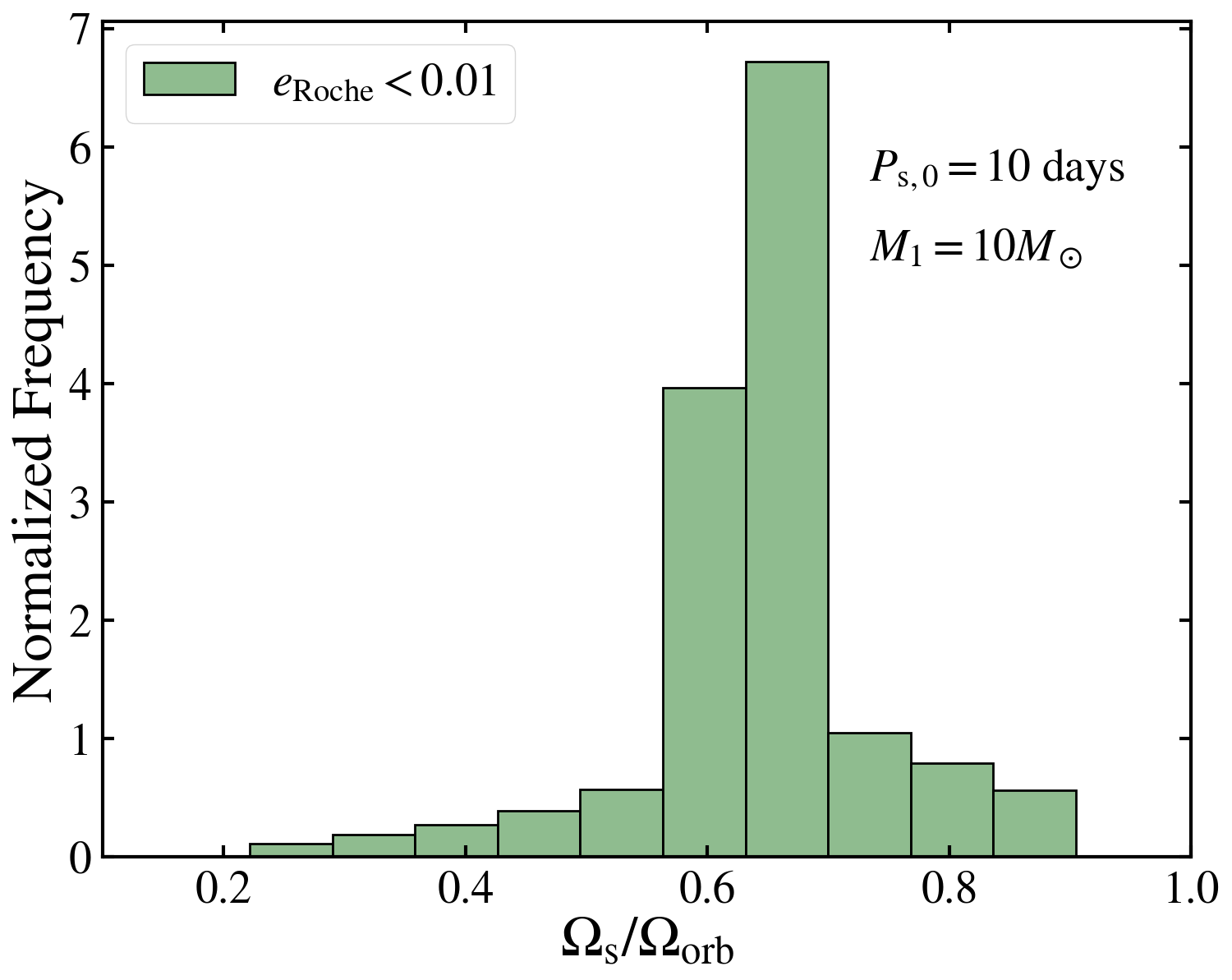}
				\includegraphics[width = 3.25 in]{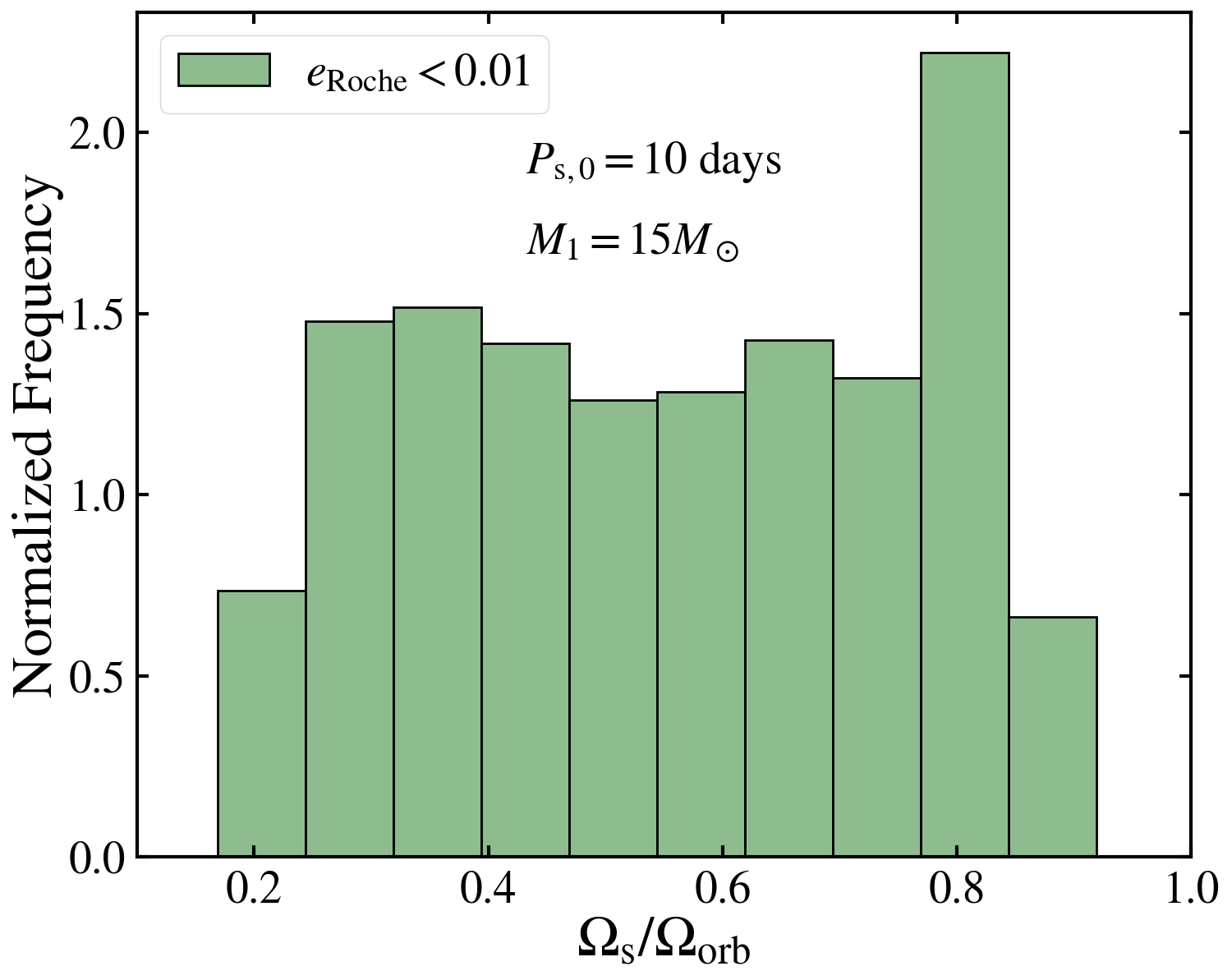}
				\caption{A histogram of the rotation rate as a fraction of the orbital frequency for systems with an eccentricity at the Roche radius of $e_{\rm Roche}<0.01$. The left and right panels show the rotation rate for the $M_1 = 10~M_\odot$ and $M_1=15~M_\odot$ stellar models respectively. The top and bottom panels show results for an initial rotation period of 1 day and 10 days respectively.}
				\label{fig:HistogramContact}
			\end{center}
		\end{figure*}

	In Fig.~\ref{fig:HistogramContact}, we provide the distribution of rotation rates for binaries that have circularized ($e_{\rm Roche} <0.01$) before the primary reaches the Roche radius. This distribution was obtained in the same way as for $e_{\rm Roche}$ but by using results for $\Omega_{\rm s}$ rather than the orbital eccentricity. In the top row, we assume an initial stellar rotation period of $P_{\rm s,0} = 1$ day when the star leaves the main sequence. In the bottom row, $P_{\rm s,0} = 10$ days. Note that almost all systems are rotating subsynchronously (i.e. $\Omega_{\rm s}/\Omega_{\rm orb} < 1$) at $r_{\rm p} = r_{\rm Roche}$. This is true regardless of the initial rotation period of the star. The subsynchronous rotation rates can be explained by competition between the tidal torque spinning up the star, and stellar expansion spinning the star down. Our results suggest that many giant stars may be rotating more slowly than expected at the onset of binary mass transfer.
	
%		\begin{figure*}
%			\begin{center}
%				\includegraphics[width = 3.25 in]{ContactCirc_M15}
%				\includegraphics[width = 3.25 in]{RocheCirc_M15}
%				\caption{15 Msol star Left: Contact, Right: Roche radius}
%			\end{center}\label{fig:FinalEccM15}
%		\end{figure*}
	
\section{Discussion}\label{sec:Discussion}

Next we discuss some caveats associated with our analysis and the implications of our results for common envelope phases and their outcomes. 

\subsection{Possible Limitations}\label{sec:limitations}

Throughout this paper, we focus on dissipation through turbulent viscosity in the convective envelope. In particular we use a linear treatment, and consider only the dissipation of the tidally excited $l=2$ fundamental mode. This formalism is most accurate when the binary separation is larger than a few times the stellar radius and when the primary star is slowly rotating. As the binary separation decreases (and becomes less than $\sim 2 R_1$), higher degree fundamental modes contribute more and more significantly to the tidal response of the star. When the binary orbital period is resonant with a fundamental mode period, these higher degree oscillations can dominate the tidal response of the star \citep{MacLeod19}, resulting in enhanced tidal dissipation. Including f-modes with higher $l$ in our calculation would likely decrease the tidal circularization time at small orbital separations.

We have not included the contribution from inertial modes in our study of tidal dissipation in a giant star. Inertial modes are restored by the Coriolis force, and can be excited when the tidal forcing frequency $\omega$ is less than $2\Omega_{\rm s}$ \cite[e.g.][]{Ogilvie07}. In a circular orbit, the forcing frequency is $2\Omega - 2\Omega_{\rm s}$. We expect that giant stars will be slowly rotating after expanding, and for many systems, inertial waves will not contribute to tidal dissipation. In cases where the tidal synchronization time is short enough that the primary is spun-up close to $\Omega_{\rm s} \sim \Omega$, the presence of inertial waves may result in enhanced dissipation.

We have also ignored possible tidal dissipation within the radiative core. This is important when ingoing gravity waves excited at the radiative-convective boundary can damp within a group velocity travel time to the center and back. In the `traveling wave' regime, the dissipation rate is given by the luminosity of the ingoing wave. The calculation of the tidal torque from gravity waves in the radiative core was first carried out by \cite{Goodman98} and later re-expressed in \citet{Kushnir17}. The form of the tidal torque in \cite{Kushnir17} reveals a strong dependence on the radius of the core/envelope boundary, $r_{\rm c}$. The strength of the torque scales as $(r_{\rm c}/R_1)^{9}$, where $R_1$ is the radius of the star. In the massive giant stars that we are considering, $r_{\rm c}/R_1\sim$ a few percent after the convective envelope has fully developed. In consequence, the tidal dissipation rate from damped internal gravity waves is negligible compared to dissipation in the envelope. 

Radiative damping can also lead to significant dissipation when gravity waves excited at the radiative-core boundary are damped before they can travel to the centre of the star. \citet{Ivanov13} derived a criterion for determining when a star is in this `Moderately Large Damping' regime. To evaluate whether this regime could apply to our stellar models, we used the stellar oscillation code GYRE \citep{Townsend13} to calculate the eigenfrequencies $\omega_\alpha$ and profiles of a few g-modes for each model. We then estimated the radiative damping rate $\gamma_{\rm rad}$ of the the modes via equation (18) of \citet{Fuller12a} \citep[see also][]{Burkart12}. We found that, for both the $10 M_\odot$ and $15 M_\odot$ stellar models, $\gamma_{\rm rad} > \omega_\alpha$, even for higher-frequency g-modes that approach the dynamical frequency of the star. This failure of the quasi-adiabatic approximation ($\gamma_{\rm rad} \ll \omega_\alpha$) implies that g-modes would not be excited in these evolved, massive stars. In any case, because of the small size of the radiative core, we expect these g-modes to have a negligible contribution to the net tidal dissipation in the star.

Our study did not include the effect of orbital decay due to hydrodynamical drag on the neutron star as it moves through the stellar wind. In some cases, this orbital decay could be comparable to the orbital expansion rate due to wind-driven mass loss, given in equation~(\ref{eq:adot_wind}). For a circular orbit with Keplerian velocity $v_{\rm k}$, we estimate the drag force as $F \sim (\uppi/2)\rho_{\rm w} R_{\rm BH}^2 v_k^2$, where $\rho_{\rm w} = \dot{M}_1/(4 \uppi a^2)$ is the wind density and $R_{\rm BH}$ is the Bondi-Hoyle radius. This gives an estimate for the ratio of the decay rates due to hydrodynamical drag and due to the stellar wind of
\begin{equation}
f\sim \frac{1}{4}\left(\frac{R_{\rm BH}}{a}\right)^2\frac{v_k}{v_{\rm w}}\frac{M_1}{M_2},
\end{equation}
Assuming that the wind velocity $v_{\rm w}$ is a few 10s of kms, $f$ can be of order unity for massive star orbiting a neutron star at a few AU. The stellar mass loss rate, $\dot{M}_1$, determines whether hydrodynamical drag affects the orbital evolution of the neutron star. For the $10M_\odot$ stellar model, the minimum timescale for mass loss during core helium burning is $M_1/\dot{M}_1 \approx 3\times 10^7$~yr. In this case, neither stellar mass loss nor hydrodynamical drag have a large effect on the binary orbital evolution. However, for the $15M_\odot$ stellar model, the mass loss timescale reaches $M_1/\dot{M}_1 \approx 3\times 10^6$~yr, which suggests that hydrodynamical drag may significantly circularize the orbit of binaries that do not reach RLO when the stellar radius first expands.

Lastly, when the binary has reached RLO, the magnitude of the tidal distortion can be of order $ \sim 10 \%$ the size of the object (depending on the eccentricity), and we assume that mass loss takes over as the primary driver of subsequent orbital evolution.  Under these conditions, our formalism is not expected to remain accurate. 
However, tidal dissipation may continue to play a large role in circularising the binary. As the binary separation continues to decrease, the interaction between the two bodies will only become much stronger. For instance, in Fig.~\ref{fig:3Cases}, when we extend our calculation beyond the Roche limit for the small separation case (the black line), the orbit fully circularizes before $r_{\rm p} = R_1$. Eventually, the tidal distortion of the primary can no longer be treated as a perturbation, and tidally excited oscillations will likely damp non-linearly through, e.g., coupling between multiple oscillation modes. In cases where the donor star is not synchronously rotating, the strong distortion of the primary may have an interesting effect on mass transfer \citep[e.g. as seen in the simulations of oscillating stars in][]{MacLeod19}. 

\subsection{Implications for Common Envelope Phases}

Traditionally, we have imagined that common envelope phases largely result from unstable RLO in circular orbits \citep[e.g. for recent examples of simulations of the onset of common envelope phases in circular orbits, see][]{2008ApJ...672L..41R,2012ApJ...746...74R,2014ApJ...786...39N,2016MNRAS.460.3992N,2017MNRAS.464.4028I,2018ApJ...863....5M,2018ApJ...868..136M,2020ApJ...893..106M,2020ApJ...895...29M}. This assumption is likely motivated by the fact that many near-contact binaries are largely circular, and by the simplicity associated with the circular-orbit limit. Further, circularized orbits are likely in common envelope systems involving lower-mass giants \citep[e.g.][]{ 2013A&ARv..21...59I,2014ApJ...786...39N}, because the stellar evolution timescale is slow compared to the tidal circularization timescale, which scales relatively weakly with stellar mass (as indicated, for example, by equation~\eqref{eq:gamma_est}). The population synthesis models of \citet{2020arXiv200109829V} and the more detailed analysis of this work, however, indicate that in many systems involving massive donors with lower-mass companions, the orbit does not circularize prior to the onset of mass exchange. 

This conclusion has several implications for the subsequent evolution of the system. Critically, the dynamics of eccentric mass transfer in binary systems are an area of active study, including analytic predictions \citep{ 2007ApJ...667.1170S, 2009ApJ...702.1387S,2010ApJ...724..546S,2016ApJ...825...70D,2016ApJ...825...71D,2017ApJ...844...12D,2019ApJ...872..119H} and gas dynamical modeling \citep{2009MNRAS.395.1127C,2011ApJ...726...66L,2011ApJ...726...67L,2016MNRAS.455.3511S,2017MNRAS.467.3556B}. In particular, the mass lost from the donor during each peripase passage, and the angular momentum imparted to it determine the orbital evolution. It remains unclear whether systems are expected to circularize following the onset of mass removal from the donor, or remain eccentric as the interaction proceeds. Because of these modifications in mass and angular momentum exchange, it is very likely that the binary eccentricity will affect the subsequent stability of mass exchange and thus determine whether or not the binary will undergo a common envelope phase. 

The outcomes of common envelope phases may depend on the initial dynamics and eccentricity of the orbit in ways that are not yet clear. One hydrodynamic study of a common envelope phase with an eccentric onset has been carried out by \citet{2016MNRAS.455.3511S}. In these models, the binary typically consisted of a $3.05M_\odot$ asymptotic giant branch donor interacting with a $1.7M_\odot$ point-mass accretor, implying a mass ratio of approximately 0.55. The periapse was chosen such that the donor lost on the order of 10\% of its own mass in the first passage, which had eccentricities varying between 0.33 and 0.7 across the models considered. \citet{2016MNRAS.455.3511S} found that after 2-3 passages, the donor's envelope had inflated sufficiently and the eccentricity had decreased such that the accretor particles did not leave the donor's envelope. After this point, \citet{2016MNRAS.455.3511S} argued that the ensuing common envelope phase carries qualitative similarity to models of common envelope phases that are initialized in circular orbits. Finally, \citet{2016MNRAS.455.3511S} noted that they are not able to estimate how eccentricity affects the final separation of their common envelope models, because it is limited by spatial resolution in their simulations. 

It is worth noting that many simulations of common envelope phases that are initialized with circular orbits develop eccentricity as the objects plunge together \citep[for some recent examples, see][]{2016ApJ...816L...9O,2018MNRAS.480.1898C,2019MNRAS.486.5809P}. This may be due to artificial initial conditions in which the accretor is placed in a non-equilibrium configuration near the surface of the donor \citep[see Figure 4 of][for example]{2017MNRAS.464.4028I}. However, the subsequent behavior that is observed is that this eccentricity is either maintained or is slow to dissipate as the orbital separation tightens \citep[e.g. Figure 1 of][]{2016ApJ...816L...9O}. At the simplest level, the details of these dynamics have to do with the relative drag forces at periapse and apoapse, which depend on the density profile of the common envelope as well as the relative velocities determined by the eccentric orbit. 

These previous simulation results are suggestive that some eccentricity may be maintained by a common envelope system as it coalesces. In subsequent phases, as the orbital tightening slows, previous modeling suggests that the remaining eccentricity may be reduced. The implications of this eccentricity for the orbital inspiral dynamics may lie largely in the way that the eccentric orbit impinges upon and modifies the structure of the common envelope. This will be particularly true if an accretion disk forms around the accretor following each periapse passage \citep{2016MNRAS.455.3511S}, which might supply energetic feedback to the surroundings via a disk wind or jet \citep{2020arXiv200314096S}. The extreme case of this scenerio, outlined by \citet{2020arXiv200314096S} is that the envelope is continuously removed and the objects never fully immerse into a shared envelope. In some rare systems with very high eccentricity, the neutron star may even enter and escape a common envelope episode \citep{Gilkis19}. The neutron star accretion of envelope material and subsequent launching of jets could result in a repeating transient, and ultimately a `common envelope jets supernova' \citep{Soker18}, a proposed explanation for supernova iPTF14hls. Even in less extreme scenarios, as the objects spiral through a modified envelope structure we might expect variations in the resultant dynamics and common envelope outcome \citep[e.g. as discussed in the context of drag forces by][]{2019MNRAS.490.3727C}.

The answers to many of these questions surrounding the importance of eccentricity in common envelope phase dynamics await future, more systematic study. Our present results indicate, however, that the eccentric onset of common envelope phases involving massive-star donors may be the norm, rather than a special case. 

Our results may seem in tension with the population synthesis of \citet{Hurley02} which coupled weak tidal friction and binary evolution, and found that most binaries will tidally circularize before RLO. However,our investigation differs from theirs in that we focus on the case of massive-stars that reach the Roche radius during a phase of rapid radius expansion.

\subsection{Implications for the Formation of Gravitational-Wave Mergers}

The LIGO and VIRGO network of gravitational wave detectors have demonstrated that long-posited mergers of binary black holes and neutron stars occur with remarkable frequency in the local universe \citep{2017PhRvL.119p1101A,2019PhRvX...9c1040A}. While there are many conclusions to draw from this new abundance of empirical data, one of the remaining uncertainties is how these populations of stellar remnants are assembled into tight orbits. For both black hole and, especially, neutron star systems, common envelope phases are one of the key possible assembly channels \citep[as discussed by, e.g.][]{ 1973A&A....25..387V,1976IAUS...73...75P,1984ApJ...277..355W,1984Natur.309..235V,1993PASP..105.1373I,1994PASAu..11...82T,2002ApJ...572..407B,2007PhR...442...75K,2008ApJS..174..223B,2012ApJ...759...52D,2013A&ARv..21...59I}. Black holes have other plausible assembly channels, including dynamical interactions in hierarchical multiple stars or dense clusters \citep[e.g.][]{2011ApJ...741...82T,2013MNRAS.435..943P,2014MNRAS.439.1079A,2014ApJ...784...71S,2015ApJ...800....9M,2016ApJ...830L..18B,2016ApJ...824L...8R,2016PhRvD..93h4029R,2016ApJ...816...65A}, isolated evolution in initially-close binaries \citep{2016MNRAS.458.2634M,2016MNRAS.460.3545D,Silsbee17,Liu18,Hamers18,Liu19a,Liu19b,Liu19c}, or a combination of cluster dynamics and binary stellar evolution \citep[e.g.][]{2014ApJ...794....7M,2014MNRAS.441.3703Z,2018MNRAS.474.2959G,2018MNRAS.479.4391M,2019MNRAS.487....2M,2019MNRAS.485..889S,2019MNRAS.487.2947D}. 

Enriching our understanding of the gravitational-wave source population requires more detailed study of their assembly channels.  Recently, \citet{2018A&A...615A..91B} have compared estimated rates for double neutron star systems from various channels \citep[also see the excellent review double neutron star assembly of][]{2017ApJ...846..170T}.  Population models of binary neutron star mergers due to common envelope phases have trouble reproducing the currently estimated merger rate \citep{2017PhRvL.119p1101A,2019PhRvX...9c1040A} on the basis of the GW170817 merger \citep{2018A&A...615A..91B,2018MNRAS.481.4009V,2018MNRAS.479.4391M,2019ApJ...880L...8A,2020ApJ...892L...9A}. This alone indicates that more detailed work is needed to understand these sources evolutionary history. 

In our modeling of tidal evolution, we have focused on pre-common envelope binary parameter combinations relevant to the formation of double neutron star systems, as studied by \citet{2020arXiv200109829V}. The interaction of giant stars with $1.4M_\odot$ companion is representative of the evolutionary stage following the first supernova. Such binaries inevitably have high eccentricities because of the large supernova kicks \citep[a few 100s of km/s or more; see, e.g.,][]{Cordes98,Lai01}  after the formation of the first neutron star. The ensuing common envelope phase is thought to be crucial to the assembly of double neutron star systems into tight orbits (for a helpful evolutionary schematic see the cartoons of \citet{2020arXiv200109829V}'s or \citet{2018A&A...615A..91B}'s Figure 1). Our findings demonstrate that the majority of these systems will not circularize prior to the onset of these common envelope phases.

The expected impact of pre-common envelope eccentricity on these common envelope phases remains uncertain, as discussed above. It is, however, clear that these phases merit further study, with attention to the ways in which pre-common envelope eccentricity may affect the post-common envelope system that emerges. These results could then be applied to next-generation binary population models to study the statistical impacts on the merging compact object rate and population.

\section{Summary \& Conclusion}\label{sec:Summary}

In this paper, we have studied how coupled stellar evolution and tidal dissipation affect the orbital parameters of binaries with a primary star that has left the main sequence and a compact star companion. These systems may undergo subsequent mass transfer or common envelope phases, and eventually form compact object binaries that produce a gravitational wave signal as they merge.
Although we define a more general methodology, we have focused our analysis on systems similar to those that undergo common envelope phases leading to the formation of merging double neutron stars \citep{2020arXiv200109829V}.

Our analysis was performed by coupling models of evolving giant star primaries with the tidal evolution of the eccentric binary, based on the the theory recently developed in VL20. We generated two stellar models with MESA (one $10 M_\odot$ and one $15 M_\odot$) to describe the stellar radius and viscous damping rates as functions of time as the primary star develops a convective envelope (see Section~\ref{sec:Models} and Figs.~\ref{fig:StellarEvolution} and \ref{fig:Zoom}). We considered a neutron star companion of mass $1.4 M_\odot$. We then used the orbital evolution equations~(\ref{eq:Simple_adot} - \ref{eq:Simple_edot}) and the time-dependent stellar structure to calculate the binary $a$ and $e$ and the stellar spin rate $\Omega_{\rm s}$ on the way to RLO. Our tidal evolution model is accurate even for close pericentre distances and highly eccentric orbits, going beond the commonly used weak friction tidal model that underestimates the strength of dissipation (VL20 and Fig. 3). 
Our main findings are:
\begin{enumerate}[i), leftmargin=3pt,itemsep=1ex]
\item We identify  three main outcomes of the coupled stellar and orbital evolution, which are highlighted in Fig.~\ref{fig:3Cases}:
      \begin{enumerate}[1), leftmargin=\parindent]
      		 \item The binary does not circularize before the primary overflows its Roche lobe;
    		 \item The binary completely circularizes before the onset of RLO;
    		 \item The binary is too wide to undergo RLO.
\end{enumerate}
The initial properties of the binary orbital configuration largely determine which of these outcomes arises.

\item We find that orbital eccentricity at RLO depends very sensitively on the initial pericentre distance, $r_{\rm p,0}$ and eccentricity $e_0$, before the primary leaves the main sequence. For the $10 M_\odot$  ($15 M_\odot$) primaries interacting with $1.4M_\odot$  companions, systems with $r_{\rm p,0} \lesssim 3$~AU (6~AU) retain their initial eccentricity at the onset of RLO (Section~\ref{sec:Outcome} and Figs. \ref{fig:FinalEcc_M10} and \ref{fig:FinalEcc_M15}).

\item For $10M_\odot$ ($15M_\odot$) plus $1.4M_\odot$ binaries that eventually overflow their Roche lobes, given an initially thermal eccentricity distribution and a log-uniform semimajor axis distribution, 75\% (63\%) of the systems overflow their Roche lobes before a convective envelope develops at eccentricities similar to their initial eccentricity values, 12\% (33\%) develop a convective envelope but overflow their Roche lobes at $e>0.1$, and 13\% (4\%) develop a convective envelope and circularize prior to RLO at $e<0.1$ (see Figs.~\ref{fig:FinalEcc_M10}-\ref{fig:HistogramRoche}).

\item For systems that do not develop a convective envelope prior to RLO, tidal dissipation has little effect on the stellar spin, and these donors are likely to spin at a rate similar to the initial value at the end of the main sequence (Figs. \ref{fig:FinalEcc_M10} and \ref{fig:FinalEcc_M15}). For systems that do develop a convective envelope prior to RLO, the stellar spin slows significantly due to structural changes in the star. In general, the spin rate at RLO correlates with the degree of tidal circularization. For systems that do not circularize, tides do not have a significant effect on the stellar spin rate; these systems rotate significantly subsynchronously at RLO (Figs. \ref{fig:FinalEcc_M10} and \ref{fig:FinalEcc_M15}). In binaries that do tidally circularize, the stellar rotation period approaches synchronicity, but remains longer than the orbital period, typically by a factor of a few because tidal dissipation and stellar evolution are acting at similar rates (Figs. \ref{fig:FinalEcc_M10},  \ref{fig:FinalEcc_M15},  and \ref{fig:HistogramContact}).  
\end{enumerate}

Our results suggest that a detailed analysis of the interaction of tides and stellar evolution is needed in systems containing evolving massive-stars because the evolutionary (for example radius growth) timescale is similar to the tidal dissipation timescale. 
Our modeling further indicates the critical importance of the phases just prior to RLO, when tides are at their strongest. In this regime our tidal evolution model (VL20) predicts dissipation rates up to $10^2$ times greater than that of the more commonly applied weak friction model.

Our analysis reveals that it is likely that many systems that undergo common envelope phases involving massive donor stars do so with initially eccentric orbits at the time of RLO \citep{2020arXiv200109829V}. These eccentric interactions remain poorly understood and deserve further study.

\section*{Acknowledgements} 

We are grateful to I. Mandel and A. Vigna-G\'omez for many helpful discussions about the effects of tides on binary populations. We thank P. Ivanov, for a thoughtful review that improved the manuscript. We thank S. E. de Mink for providing the MESA inlists that we adapted to create the stellar models used in our study.

This work was supported
by NSF grants
1909203 and AST-1715246 as well as NASA grant 80NSSC19K0444.

\section*{Data Availability Statement}
The MESA and GYRE inlists used to create and analyze our stellar models are available from the corresponding author upon request. The MESA inlists will be posted on the MESA marketplace (http://cococubed.asu.edu/mesa\_market/inlists.html).

\bibliographystyle{mnras}	
\bibliography{References,ce1,ce2}

\bsp
\label{lastpage}

\end{document}